\documentclass[twoside,journal]{IEEEtran}
\IEEEoverridecommandlockouts
\renewcommand\IEEEkeywordsname{Index Terms}
\ifCLASSINFOpdf
\else
\fi
\usepackage{xcolor,soul,framed} 
\colorlet{shadecolor}{yellow}
\usepackage[pdftex]{graphicx}
\graphicspath{{../pdf/}{../jpeg/}}
\DeclareGraphicsExtensions{.pdf,.jpeg,.png}
\usepackage{cite}
\usepackage{amsmath,amssymb,amsfonts}
\usepackage{algpseudocode}
\PassOptionsToPackage{ruled, linesnumbered}{algorithm2e}
\usepackage{algorithm2e}
\SetKw{Break}{break out of the loop}
\usepackage{graphicx}
\usepackage{textcomp}
\usepackage{comment}
\usepackage{units}
\usepackage{url}
\usepackage{geometry}
\usepackage{bbding}
\usepackage[caption=false,font=footnotesize]{subfig}
\usepackage{mathrsfs}
\usepackage{xcolor}
\definecolor{myblue}{rgb}{0.0, 0.5, 1.0}
\definecolor{myred}{rgb}{1.0, 0.13, 0.32}
\definecolor{mygreen}{rgb}{0.31, 0.68, 0.07}
\definecolor{grn}{rgb}{0.0, 0.5, 0.0}
\usepackage[pdftex]{graphicx}
\DeclareGraphicsExtensions{.pdf,.jpeg,.png}
\def\BibTeX{{\rm B\kern-.05em{\sc i\kern-.025em b}\kern-.08em
    T\kern-.1667em\lower.7ex\hbox{E}\kern-.125emX}}

\usepackage{pgfplots}
\usepackage{pgfplotstable}
\usepackage{tikz}
\usetikzlibrary{calc}

\usepackage{hyperref}
\hypersetup{
	colorlinks=true,
	linkcolor=blue,
	filecolor=magenta,      
	urlcolor=blue,
	citecolor=blue,
}

\usepackage{balance}
\allowdisplaybreaks

\begin{document}
   \title{\huge Exploiting Structural Flexibility in SIM-Enabled Communications: From Adaptive Inter-Layer Spacing to Fully Morphable Layers} 
    	\newgeometry {top=25.4mm,left=19.1mm, right= 19.1mm,bottom =19.1mm}%
\author{Ahmed Magbool,~\IEEEmembership{Graduate Student Member,~IEEE,} 
Vaibhav Kumar,~\IEEEmembership{Member,~IEEE,} 
\\ Marco Di Renzo,~\IEEEmembership{Fellow,~IEEE,} and Mark F. Flanagan,~\IEEEmembership{Senior Member,~IEEE}\vspace{-0.8cm}\thanks{Ahmed Magbool and Mark F. Flanagan are with the School of Electrical and Electronic Engineering, University College Dublin, D04 V1W8, Ireland (e-mail: ahmed.magbool@ucdconnect.ie, mark.flanagan@ieee.org). \par
Vaibhav Kumar is with Engineering Division, New York University Abu Dhabi (NYUAD), Abu Dhabi 129188, UAE (e-mail: vaibhav.kumar@ieee.org).\par 
Marco Di Renzo is with Universit\'e Paris-Saclay, CNRS, CentraleSup\'elec, Laboratoire des Signaux et Syst\`emes, 3 Rue Joliot-Curie, 91192 Gif-sur-Yvette, France (marco.di-renzo@universite-paris-saclay.fr), and with King's College London, Centre for Telecommunications Research -- Department of Engineering, WC2R 2LS London, United Kingdom (marco.di\_renzo@kcl.ac.uk). \par
The work of A. Magbool and M. F. Flanagan was supported by Research Ireland under Grant Number 13/RC/2077\_P2 and under Grant Number 24/FFP-P/12895. \par
The work of M. Di Renzo was supported in part by the European Union through the Horizon Europe project COVER under grant agreement number 101086228, the Horizon Europe project UNITE under grant agreement number 101129618, the Horizon Europe project INSTINCT under grant agreement number 101139161, and the Horizon Europe project TWIN6G under grant agreement number 101182794, as well as by the Agence Nationale de la Recherche (ANR) through the France 2030 project ANR-PEPR Networks of the Future under grant agreement NF-YACARI 22-PEFT-0005, and by the CHIST-ERA project PASSIONATE under grant agreements CHIST-ERA-22-WAI-04 and ANR-23-CHR4-0003-01. Also, the work of M. Di Renzo was supported in part by the Engineering and Physical Sciences Research Council (EPSRC), part of UK Research and Innovation, and the UK Department of Science, Innovation and Technology (CHEDDAR Telecom Hub, grants reference EP/X040518/1 and EP/Y037421/1, and HASC Telecom Hub, grants reference EP/X040569/1).  \par
A preliminary version of this paper is to be presented at the
IEEE International Conference on Communications (ICC 2026)~\cite{2025_Magbool1}.} }


\maketitle
\begin{abstract}
Stacked intelligent metasurfaces (SIMs) have recently emerged as a promising metasurface-based physical-layer paradigm for wireless communications, enabling wave-domain signal processing through multiple cascaded metasurface layers. However, conventional SIM designs rely on rigid planar layers with fixed interlayer spacing, which constrain the propagation geometry and can lead to performance saturation as the number of layers increases. This paper investigates the potential of introducing structural flexibility into SIM-enabled communication systems. Specifically, we consider two flexible SIM architectures: distance-adaptive SIM (DSIM), where interlayer distances are optimized, and stacked flexible intelligent metasurface (SFIM), where each metasurface layer is fully morphable. We jointly design the meta-atom positions and responses together with the transmit beamformer to maximize the system sum rate under per-user rate, quantization, morphing, and interlayer distance constraints. An alternating optimization framework combining gradient projection, penalty-based method, and successive convex approximation is developed to address the resulting non-convex problems. Perturbation analysis reveals that the flexibility gains of both DSIM and SFIM scale approximately linearly with the morphing range, with SFIM exhibiting a faster growth rate. Simulation results demonstrate that flexible SIM designs mitigate performance saturation with increasing layers and achieve significant transmit power savings compared to rigid SIMs.
\end{abstract}

\IEEEpeerreviewmaketitle
\vspace{-0.4cm}
\section{Introduction} \label{sec:intro}

Massive multiple-input multiple-output (mMIMO) is a key technology for fifth-generation (5G) wireless networks due to its ability to provide highly directional beamforming and exploit spatial diversity through a large number of antenna elements. However, realizing these gains typically requires large-scale phased antenna arrays with finely quantized phase and amplitude control. Such high-resolution hardware architectures substantially increase system cost, circuit complexity, and power consumption, posing practical challenges for scalable and energy-efficient deployment~\cite{2014_Larsson,2014_Lu}, thereby motivating alternative physical-layer paradigms that reduce reliance on high-resolution active antenna arrays.

To overcome these challenges, metasurfaces have emerged as a compelling candidate for future sixth-generation (6G) networks. Electromagnetic metasurfaces are programmable structures composed of artificial electromagnetic materials that exhibit functionalities not achievable with non-reconfigurable surfaces, leading to enhanced system performance~\cite{2019_Renzo}. By densely packing sub-wavelength scattering elements, metasurfaces enable fine-grained control over the wireless propagation environment while relying on low-power hardware~\cite{2015_Achouri}. Depending on their configuration, intelligent metasurfacess can operate as transmitting, receiving, or reflecting nodes, thereby enabling programmable control of the wireless propagation environment~\cite{2020_Huang,2025_Magbool}.

Building on this, the concept of stacked intelligent metasurfaces (SIMs) has recently emerged, where multiple metasurface layers are cascaded to perform signal manipulation directly in the electromagnetic domain~\cite{2025_Liu}. Unlike traditional mMIMO transceiver designs, SIM-based architectures offer a more energy- and cost-efficient solution by shifting part of the signal processing from power-intensive digital baseband units to nearly-passive metasurface layers. The application of SIMs in multi-user communication was initially explored in~\cite{2023_An}, which showed that SIM-enabled systems can markedly reduce the precoding latency associated with digital beamforming. Subsequent work in~\cite{2023_An1} demonstrated that SIM-based designs can achieve capacity gains of up to 150\% compared to conventional MIMO and reconfigurable intelligent surface (RIS)-aided systems. In~\cite{2025_Papazafeiropoulos}, a dual-SIM architecture was proposed, with one SIM deployed at the transmitter and another near the users, leveraging statistical channel state information (CSI) to reduce the need for frequent re-optimization within each coherence interval. A related two-SIM configuration was studied in~\cite{2024_Papazafeiropoulos}, where the achievable rate was optimized for systems equipped with SIMs at both the transmitting and receiving ends. Beyond these configurations, SIMs have been investigated in a variety of contexts, including physical-layer security~\cite{2024_Niu,2025_Kavianinia}, cell-free MIMO systems~\cite{2024_Li,2025_Hu}, wireless sensing~\cite{2025_Teng}, and metasurfaces with active or amplifying elements~\cite{2025_Darsena}. Despite these advantages, the existing SIM designs assume rigid planar layers with fixed inter-layer spacing, which constrains the propagation geometry across layers and can lead to performance saturation as the number of layers increases.

Recently, flexible transceiver architectures, such as movable antennas~\cite{2026_Zhu}, fluid antennas~\cite{2025_New}, and pinching antennas~\cite{2025_Liu1}, have been investigated to exploit spatial reconfigurability and introduce an additional degree of control in wireless communication systems~\cite{2025_You}. In addition, flexible intelligent metasurfaces (FIMs) have been proposed as deformable and reconfigurable surfaces whose electromagnetic response and physical shape can be jointly adjusted to adapt wireless propagation in real time~\cite{2025_An3}. Compared to conventional rigid metasurfaces, FIMs offer enhanced adaptability, allowing the surface to preserve favorable propagation conditions across changing environments~\cite{Kumar_FIM_Letters,2025_Mursia}. In~\cite{2025_An}, it was shown that deploying a transmit FIM at the base station (BS) can reduce power consumption by up to 50\% relative to rigid antenna arrays while meeting identical quality-of-service (QoS) constraints. The capacity maximization of point-to-point MIMO systems incorporating FIMs at both the transmitting and receiving ends was examined in~\cite{2025_An1}, where the results indicated that FIM-based architectures can achieve up to a twofold increase in channel capacity compared to systems using rigid arrays. Furthermore, the application of FIMs as reflective surfaces was studied in~\cite{2025_Hu3}, where optimizing the channel gain revealed substantial performance improvements over traditional rigid RISs, highlighting the benefits of surface flexibility. {However, these studies on FIMs have focused on \emph{single-layer} or \emph{dual-surface} configurations, and their integration into stacked SIM architectures remains largely unexplored.}

In this paper, we propose two approaches to incorporate flexibility into the SIM architecture: (i) optimizing the inter-layer distances between the SIM layers, and (ii) employing FIMs as layers within the SIM architecture. We refer to these architectures as distance-adaptive SIM (DSIM) and stacked FIM (SFIM), respectively. In particular, the main contributions of this paper can be summarized as follows:
\begin{itemize}
    \item {We propose two flexible SIM architectures: a distance-adaptive SIM (DSIM), where inter-layer spacing is optimized while layers remain rigid, and a stacked flexible intelligent metasurface (SFIM), where each SIM layer is fully morphable.}
    \item We formulate an optimization problem for each of these proposed systems. In both cases, we maximize the system sum rate subject to a minimum rate constraint for each user, a power budget constraint, a morphing limit constraint for each meta-atom, and a minimum inter-layer element spacing constraint. We also assume that the meta-atom responses are discrete. For the DSIM system, additional constraints are imposed to ensure rigidity of each surface.
    \item To solve the resulting nonconvex optimization problems, we develop an alternating optimization (AO) approach. In this framework, the meta-atom morphing is optimized using a gradient projection method combined with a penalty-based approach, while the transmit beamformer and the meta-atom responses are optimized using a successive convex approximation (SCA) method.
    \item We present a perturbation analysis for the single-input single-output (SISO) case, where we show that the sum rate scales approximately linearly with the morphing range, with a slope equal to the $\ell_1$-norm of the gradient of the sum rate with respect to the morphing parameter, evaluated at zero. This linear growth is also observed for the DSIM case, albeit with a lower slope. While our analysis is based on a Taylor approximation and is valid when the morphing range is not too large, simulation results show that, for larger morphing ranges, the sum rate grows more rapidly with morphing range in both the SFIM and DSIM cases.
    \item Simulation results demonstrate that both DSIM and SFIM provide significant performance gains compared to their rigid SIM (RSIM) counterpart. They further show that the two flexible SIM designs can mitigate performance saturation as the number of SIM layers increases, while also offering substantial power savings relative to RSIM.
\end{itemize}

While a relevant SIM architecture was previously presented in~\cite{2025_Niu}, that work considers the special case of a two-layer SFIM design based on the use of zero-forcing to mitigate inter-user interference. In contrast, the optimization problems formulated in this paper are more complex and incorporate practical considerations such as quantized meta-atom responses and a minimum allowable distance between meta-atoms in different layers. As a result, both the optimization framework and the perturbation analysis in this paper are fundamentally different from those presented in~\cite{2025_Niu}.

The rest of the paper is organized as follows. Section~\ref{sec:sys_model} presents the system model, including the signal and channel models. Section~\ref{sec:prob_form} formulates the optimization problems for both the SFIM and DSIM systems, and Section~\ref{sec:prop_sol} presents the proposed solution. Section~\ref{sec:per_ana} provides a perturbation analysis. Section~\ref{sec:sim} presents the simulation results and discussion, and Section~\ref{sec:conc} concludes the paper.

\begin{figure}[t]
	\centering
	\subfloat[DSIM]{
		\includegraphics[width=0.3\textwidth]{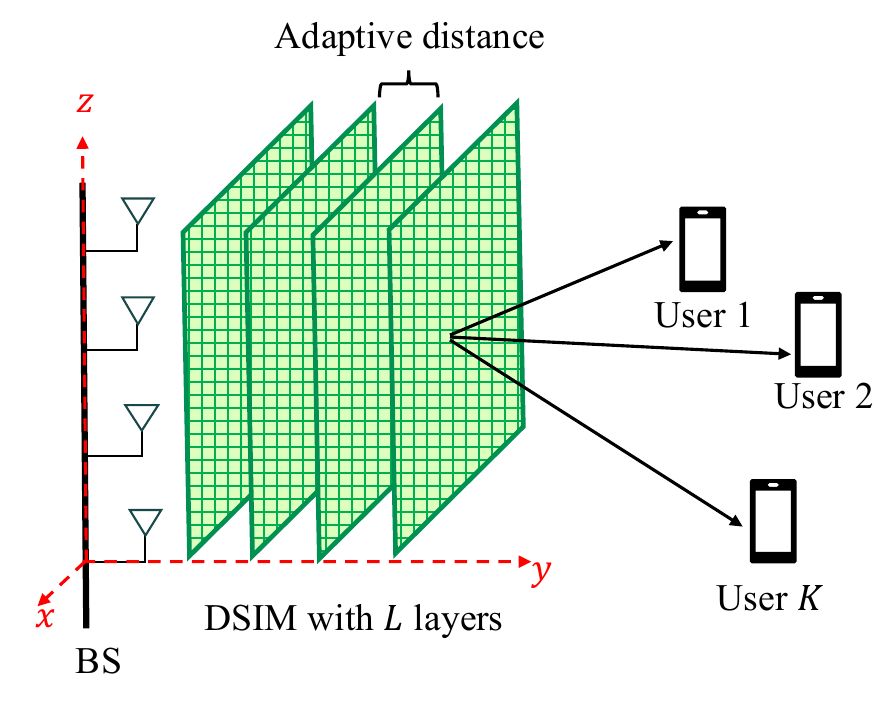}
		\label{fig:sys_model_dsim}
	}
	\hfill
	\subfloat[SFIM]{
		\includegraphics[width=0.3\textwidth]{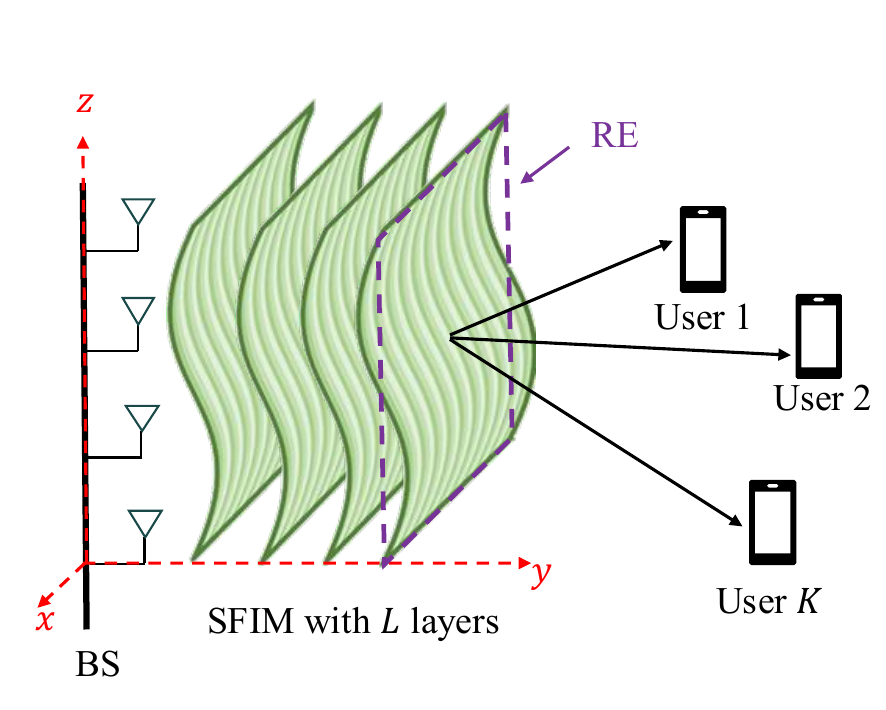}
		\label{fig:sys_model_sfim}
	}
	\caption{Proposed system model with (a) DSIM and (b) SFIM, each comprising {$M$ transmit antennas}, $L$ layers serving $K$ users.}
	\label{fig:sys_model}
\end{figure}

\textit{Notation:} Bold lowercase and uppercase letters denote vectors and matrices, respectively. $\Re \{ \cdot \}$, $\Im \{ \cdot \}$, $|\cdot|$ and $(\cdot)^*$ represent the real part, the imaginary part, the magnitude and the complex conjugate of a complex-valued matrix, respectively. $\|\mathbf{\cdot} \|_1$, $\|\mathbf{\cdot} \|_2$ and $\|\mathbf{\cdot} \|_\textsc{F}$ represent the $\ell_1$ vector norm, the Euclidean vector norm and the Frobenius matrix norm, respectively. $\mathbf{(\cdot)}^\mathsf{T}$ and $\mathbf{(\cdot)}^\mathsf{H}$ denote the matrix transpose and matrix conjugate transpose, respectively. $\mathbf{I}_a$ represents the $a \times a$ identity matrix, while $\mathbf{1}_{a\times b}$ and $\mathbf{0}_{a\times b}$ denote $a \times b$ matrices whose entries are all ones and zeros, respectively. For a diagonal matrix $\mathbf{A}$, $\text{diag}(\mathbf{A})$ represents a vector whose entries consist of the diagonal elements of $\mathbf{A}$. $\mathbb{C}$ denotes the set of complex numbers and $j = \sqrt{-1}$, while $\mathbb{B} = \{0,1 \}$. $\mathbb{E}\{ \cdot \}$ represents the expectation operator. $\mathcal{CN}(\mathbf{a},\mathbf{B})$ represents a complex Gaussian distribution with mean $\mathbf{a}$ and covariance matrix $\mathbf{B}$. $\frac{\partial \mathbf{a} (x)}{\partial x}$ is a vector representing the element-wise partial derivative of $\mathbf{a} (x)$ with respect to $x$, and $\nabla_\mathbf{a} f(\mathbf{a})$ represents the gradient of the function $f(\mathbf{a})$ with respect to $\mathbf{a}$.

\section{System Model} \label{sec:sys_model}

We consider a downlink communication system consisting of a BS and $K$ single-antenna users. The BS is equipped with $M$ antennas and an $L$-layer SIM, where each layer comprising $N$ meta-atoms. We {investigate} two variations of the system model: (i) {a} DSIM, in which 
{each metasurface layer is rigid while the inter-layer distances are adjustable}, and (ii) {an} SFIM, in which each layer of the SIM is fully {morphable}, as illustrated in Fig.~\ref{fig:sys_model}. In the following, we present the signal and channel models.

\subsection{Transmit Signal Model}
The BS transmits the following signal:
\begin{equation}
    \mathbf{x}=\sum\nolimits_{k\in \mathcal{K}} \mathbf{w}_k s_k =\mathbf{W}\mathbf{s},
\end{equation}
where $\mathcal{K} \triangleq \{1,\dots,K\}$ {denotes the set of user indices, and} $\mathbf{w}_k \in \mathbb{C}^{M \times 1}$ and $s_k \in \mathbb{C}$ denote the digital precoder and data symbol for the $k$-th user, respectively. Moreover, $\mathbf{W} \triangleq [\mathbf{w}_1,\dots,\mathbf{w}_K]$ and $\mathbf{s} \triangleq [s_{1},\dots,s_{K}]^\mathsf{T}$ with $\mathbb{E}\{\mathbf{s}\mathbf{s}^\mathrm{H}\} \sim \mathcal{CN}(\mathbf{0},\mathbf{I}_{K})$.

\subsection{Channel Model}
\begin{figure*}
  \centering
  \begin{tabular}{c c c}
    \includegraphics[width=0.5\columnwidth, height=0.42\columnwidth]{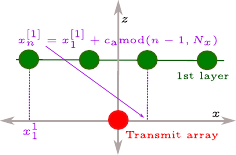} &
    \includegraphics[width=0.5\columnwidth, height=0.42\columnwidth]{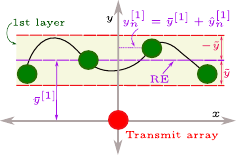} &
    \includegraphics[width=0.5\columnwidth, height=0.42\columnwidth]{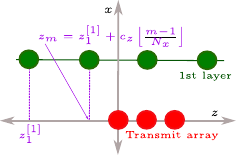} \\
    \scriptsize (a) Distances along the $x$-axis between the transmitter \vspace{-0.1cm} &
    \scriptsize (b) Distances along the $y$-axis between the transmitter &
    \scriptsize (c) Distances along the $z$-axis between the transmitter \\
    \scriptsize and the first layer. &
    \scriptsize and the first layer. &
    \scriptsize and the first layer. \\
       \includegraphics[width=0.5\columnwidth, height=0.42\columnwidth]{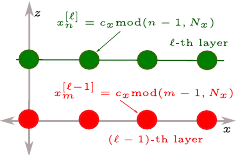} &
    \includegraphics[width=0.5\columnwidth, height=0.42\columnwidth]{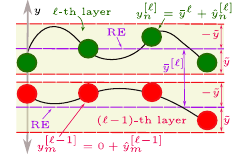} &
    \includegraphics[width=0.5\columnwidth, height=0.42\columnwidth]{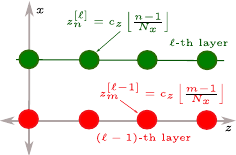} \\
    \scriptsize (d) Distances along the $x$-axis between the $(\ell-1)$th \vspace{-0.1cm} &
    \scriptsize (e) Distances along the $y$-axis between the $(\ell-1)$th  &
    \scriptsize (f) Distances along the $z$-axis between the $(\ell-1)$th  \\
    \scriptsize and the $\ell$-th layer. &
    \scriptsize and the $\ell$-th layer. &
    \scriptsize and the $\ell$-th layer. 
  \end{tabular}
    \medskip
      \vspace{-.4cm}
  \caption{Geometry of the transmitter-to-first-layer distances and the inter-layer distances.}
  \vspace{-.4cm}
  \label{fig:SC}
\end{figure*}
In this section, we present the channel model for SFIM, as it represents the more general case. At the end of the section, we describe the modifications needed to adapt the model for DSIM.

As illustrated in Fig.~\ref{fig:SC}, the transmit array is assumed to be located along the $z$-axis. The \textit{rigid equivalent} (RE) of the $\ell$-th meta-surface is then defined as a reference surface parallel to the $x$–$z$ plane, on which all meta-atoms have the same $y$-coordinate\footnote{This convention is adopted for simplicity; however, other orientations of the transmit antennas and the REs of the SFIM layers can also be employed.}. {The RE serves as a fixed geometric reference that enables the morphing-induced perturbations of the meta-atoms to be modeled explicitly while preserving a tractable channel formulation.} Furthermore, we define the distance along the $y$-axis between the transmit antennas and the RE of the first layer as $\bar{y}^{[1]}$. Similarly, we define the distance along the $y$-axis between the REs of the $(\ell-1)$-th layer and the $\ell$-th layer as $\bar{y}^{[\ell]}$ for all $\ell\in\mathcal{L}\setminus\{1\}$ with $\mathcal{L}\triangleq\{1,\dots,L\}$. We also define $[-\tilde{y},\tilde{y}]$ as the \textit{morphing range} for each meta-atom, representing the range within which the meta-atom can move along the $y$-axis. This range is assumed to be symmetric around $\bar{y}^{[\ell]}$ for all $\ell \in \mathcal{L}$.

Define $\mathbf{h}_{k}^\mathsf{T}(\hat{\mathbf{y}}) \in \mathbb{C}^{1 \times N}$ as the channel between the {$L$-th (final)} SFIM layer and the $k$-th user, $\boldsymbol{\Omega}^{[\ell]}(\hat{\mathbf{y}}) \in \mathbb{C}^{N \times N}$ as the channel between the $(\ell-1)$-th layer and the $\ell$-th layer of the SFIM for all $\ell\in\mathcal{L}\setminus\{1\}$, $\boldsymbol{\Omega}^{[1]}(\hat{\mathbf{y}}) \in \mathbb{C}^{N \times M}$ as the channel between the transmit antenna array and the first SFIM layer, $\hat{y}_{n}^{[\ell]}$ as the morphing distance of the $n$-th element in the $\ell$-th layer, and $\hat{\mathbf{y}}\triangleq[\hat{y}_{1}^{[1]},\dots,\hat{y}_{N}^{[1]},\dots,\hat{y}_{1}^{[L]},\dots,\hat{y}_{N}^{[L]}]^\mathsf{T}$. Then the cascaded channel between the transmit antenna array at the BS and the $k$-th user can be expressed as 
\begin{equation}
	\mathbf{g}_{k}^\mathsf{T}(\hat{\mathbf{y}},\boldsymbol{\phi})=\mathbf{h}_{k}^\mathsf{T}(\hat{\mathbf{y}})\boldsymbol{\Phi}^{[L]}\boldsymbol{\Omega}^{[L]}(\hat{\mathbf{y}})\dots\boldsymbol{\Phi}^{[1]}\boldsymbol{\Omega}^{[1]}(\hat{\mathbf{y}}),\label{eq:ccc}
\end{equation}
where $\boldsymbol{\Phi}^{[\ell]}\triangleq\text{diag}(\phi_{1}^{[\ell]},\dots,\phi_{N}^{[\ell]})$ represents the response of the $\ell$-th SFIM layer with $\phi_{n}^{[\ell]} \in \mathcal{Q}$ for all $\ell\in\mathcal{L}$ and for all $n\in\mathcal{N}$, $\mathcal{Q}$ is the set of the quantized meta-atom responses (i.e., a discrete set of points on the unit circle), $\mathcal{N}\triangleq\{1,\dots,N\}$, and $\boldsymbol{\phi}\triangleq[\phi_{1}^{[1]},\dots,\phi_{N}^{[1]},\dots,\phi_{1}^{[L]},\dots,\phi_{N}^{[L]}]^\mathsf{T}$.

Using the Rayleigh-Sommerfeld diffraction theory, the channel between the $m$-th transmitting meta-atom in the $(\ell-1)$-th layer (or the $m$-th transmit antenna at the BS when $\ell=1$) and the $n$-th receiving meta-atom in the $\ell$-th SFIM layer is expressed as 
\begin{multline}
	[\boldsymbol{\Omega}^{[\ell]}(\hat{\mathbf{y}})]_{n,m} = \frac{\tilde{A}^{[\ell]}\cos(\theta_{n,m}^{[\ell]})}{d_{n,m}^{[\ell]}(\hat{\mathbf{y}})} \\
	\times \bigg(\frac{1}{2\pi d_{n,m}^{[\ell]}(\hat{\mathbf{y}})}-\frac{j}{\lambda}\bigg)
	\exp\bigg(j\frac{2\pi d_{n,m}^{[\ell]}(\hat{\mathbf{y}})}{\lambda}\bigg),\label{eq:cm1}
\end{multline}
where 
\begin{equation}
	\tilde{A}^{[\ell]}\triangleq\begin{cases}
		A_{\text{a}}, & \ell=1,\\
		A_{\text{m}}, & \ell\in \mathcal{L} \setminus \{1\},
	\end{cases}
\end{equation}
with $A_{\text{a}}$ and $A_{\text{m}}$ representing the surface areas of a transmit antenna and a meta-atom, respectively, $\lambda$ denotes the carrier wavelength, $d_{n,m}^{[\ell]}(\hat{\mathbf{y}})$ is the distance between the $m$-th transmit meta-atom (or transmit antenna when $\ell=1$) in the $(\ell-1)$-th layer and the $n$-th receiving meta-atom in the $\ell$-th layer, and $\theta_{n,m}^{[\ell]}$ is the angle between the normal direction and the propagation direction between the $m$-th transmit meta-atom (or transmit antenna when $\ell=1$) in the RE of the $(\ell-1)$-th layer and the $n$-th receiving meta-atom in the RE of the $\ell$-th layer. {Next, we explicitly express the channel coefficients in terms of the morphing distances for the two relevant cases.}

\paragraph{\textit{$\boldsymbol{\Omega}^{[1]} (\hat{\mathbf{y}})$}}

Taking the first antenna at the BS as a reference point, the coordinates of the $m$-th transmit antenna are $\big[x_{m}^{[0]},\ y_{m}^{[0]},\ z_{m}^{[0]}\big]=\big[0,\ 0,\ (m-1)c_{\mathrm{a}}\big]$, where $c_{\mathrm{a}}$ is the spacing between two antenna elements.

Moreover, the {coordinates} of the $n$-th element of the first flexible metasurface are
\begin{multline}
	\big[x_{n}^{[1]},\ y_{n}^{[1]},\ z_{n}^{[1]}\big] = \big[
	x_{1}^{[1]} + c_{x} \!\!\!\!\! \mod (n-1,N_{x}), \\
	\bar{y}^{[1]} + \hat{y}_{n}^{[1]},  \ z_{1}^{[1]} + c_{z}\lfloor(n-1)/N_{x}\rfloor
	\big], \label{eq:first_layer_coord}
\end{multline}
where, $x_{1}^{[1]}$ and $z_{1}^{[1]}$ represent the $x$- and $z$-coordinates of the reference SFIM meta-atom in the RE, which is assumed to be the element located at the top-left corner, with respect to the reference antenna. Also, $c_{x}$ and $c_{z}$ denote the spacing between two adjacent meta-atoms along the $x$- and $z$-directions, respectively, and $N_{x}$ denotes the number of meta-atoms along the $x$-axis, which is assumed to be a divisor of $N$.

Then we may write 
\begin{subequations}
	\begin{align}
		d_{n,m}^{[1]}(\hat{y}_{n}^{[1]}) = & {\big\Vert x_{n}^{[1]}, \bar{y}^{[1]}+\hat{y}_{n}^{[1]}, z_{n}^{[1]}-z_{m}^{[0]}\big\Vert_2}, \label{eq:d_nm_1} \\
		\cos(\theta_{n,m}^{[1]})= & \ \big(\bar{y}^{[1]}+\hat{y}_{n}^{[1]}\big)/{d_{n,m}^{[1]}(\hat{y}_{n}^{[1]})} . \label{eq:cos_theta_nm_1}
	\end{align}
\end{subequations}
 Thus 
\begin{equation}
	[\boldsymbol{\Omega}^{[1]}(\hat{\mathbf{y}})]_{n,m}=p_{n,m}^{[1]}(\hat{y}_{n}^{[1]}) \ q_{n,m}^{[1]}(\hat{y}_{n}^{[1]}) \ r_{n,m}^{[1]}(\hat{y}_{n}^{[1]}),\label{eq:cm1_ref}
\end{equation}
where
\begin{subequations}
	\begin{align}
		p_{n,m}^{[1]}(\hat{y}_{n}^{[1]}) & \triangleq A_{\text{a}}(\bar{y}^{[1]}+\hat{y}_{n}^{[1]}) / {\big(d_{n,m}^{[1]}(\hat{y}_{n}^{[1]}) \big)^2}, \\
		q_{n,m}^{[1]}(\hat{y}_{n}^{[1]}) & \triangleq  1 / \big(2\pi {d_{n,m}^{[1]}(\hat{y}_{n}^{[1]})}\big)-j / \lambda, \\
		r_{n,m}^{[1]}(\hat{y}_{n}^{[1]}) & \triangleq\exp\big\{j 2\pi{d_{n,m}^{[1]}(\hat{y}_{n}^{[1]})} / \lambda\big\}. 
	\end{align}
\end{subequations}

\paragraph{\textit{$\boldsymbol{\Omega}^{[\ell]} (\hat{\mathbf{y}})$ for $\ell\in\mathcal{L}\setminus\{1\}$}}

Assuming that the meta-atoms of all surfaces are perfectly aligned along the $x$-$z$ plane, and taking the element in the top-left corner of the RE of the $(\ell-1)$-th surface as a reference point, {the} coordinates of the $m$-th meta-atom of the $(\ell-1)$-th layer are
\begin{multline}
	\big[x_{m}^{[\ell-1]},\ y_{m}^{[\ell-1]},\ z_{m}^{[\ell-1]}\big] = \big[
	c_{x}+\!\!\!\!\mod (m-1,N_{x}), \\ 
	{\hat{y}_{m}^{[\ell-1]}},\ z_{1}^{[1]}+c_{z}\lfloor(m-1)/N_{x}\rfloor
	\big]. \label{eq:l-1th_layer_coord}
\end{multline}
Correspondingly, the coordinates of the $n$-th meta-atom of the $\ell$-th layer are
\begin{multline}
	\big[x_{n}^{[\ell]},\ y_{n}^{[\ell]},\ z_{n}^{[\ell]}\big] = \big[
	c_{x}+\!\!\!\!\mod (n-1,N_{x}), \\ 
	\bar{y}^{[\ell]}+{\hat{y}_{n}^{[\ell]}},\ z_{1}^{[1]}+c_{z}\lfloor(n-1)/N_{x}\rfloor
	\big]. \label{eq:l-th_layer_coord}
\end{multline}

Hence, we can express
\begin{subequations}
	\begin{align}
		&  d_{n,m}^{[\ell]}(\hat{y}_{m}^{[\ell-1]},\hat{y}_{n}^{[\ell]})  = \big\Vert (x_{n}^{[\ell]}-x_{m}^{[\ell-1]}), \notag \\
		& \qquad \qquad  (\bar{y}^{[\ell]}+\hat{y}_{n}^{[\ell]}-\hat{y}_{m}^{[\ell-1]}),  (z_{n}^{[\ell]}-z_{m}^{[\ell-1]}) \big\Vert_2, \\
		&   \cos(\theta_{n,m}^{[\ell]}) = \big(\bar{y}^{[\ell]}+\hat{y}_{n}^{[\ell]}-\hat{y}_{m}^{[\ell-1]}\big) / {d_{n,m}^{[\ell]}(\hat{y}_{m}^{[\ell-1]},\hat{y}_{n}^{[\ell]})}.  \!\!
	\end{align}
\end{subequations}
Thus 
\begin{multline}
	[\boldsymbol{\Omega}^{[\ell]}(\hat{\mathbf{y}})]_{n,m} = p_{n,m}^{[\ell]}(\hat{y}_{m}^{[\ell-1]},\hat{y}_{n}^{[\ell]}) \\
	\times q_{n,m}^{[\ell]}(\hat{y}_{m}^{[\ell-1]},\hat{y}_{n}^{[\ell]}) r_{n,m}^{[\ell]}(\hat{y}_{m}^{[\ell-1]},\hat{y}_{n}^{[\ell]}),\label{eq:cml_ref}
\end{multline}
where
\begin{subequations}
	\begin{align}
		& p_{n,m}^{[\ell]}(\hat{y}_{m}^{[\ell-1]},\hat{y}_{n}^{[\ell]})  \triangleq \frac{A_{\text{m}}(\bar{y}^{[\ell]}+\hat{y}_{n}^{[\ell]}-\hat{y}_{m}^{[\ell-1]})} {{\big\{ d_{n,m}^{[\ell]}(\hat{y}_{m}^{[\ell-1]},\hat{y}_{n}^{[\ell]}) \big\}^2}}, \\
		& q_{n,m}^{[\ell]}(\hat{y}_{m}^{[\ell-1]},\hat{y}_{n}^{[\ell]}) \triangleq \frac{1}{2\pi {d_{n,m}^{[\ell]}(\hat{y}_{m}^{[\ell-1]},\hat{y}_{n}^{[\ell]})}}-\frac{j}{\lambda}, \\
		& r_{n,m}^{[\ell]}(\hat{y}_{m}^{[\ell-1]},\hat{y}_{n}^{[\ell]}) \triangleq \exp\big\{j 2\pi {d_{n,m}^{[\ell]}(\hat{y}_{m}^{[\ell-1]},\hat{y}_{n}^{[\ell]})} / \lambda \big\}. 
	\end{align}
\end{subequations}

\paragraph{\textit{$\mathbf{h}_{k}(\hat{\mathbf{y}})$}}

We assume a multipath channel between the {final} SFIM layer and the $k$-th user, {which can be expressed as}
\begin{equation}
	\mathbf{h}_{k}(\hat{\mathbf{y}})=\sum\nolimits_{i=0}^{I-1}\alpha_{i,k}\mathbf{a}(\hat{\mathbf{y}},\vartheta_{i,k},\varphi_{i,k}) \ \forall k \in \mathcal K, \label{eq:SV_CM}
\end{equation}
where $I$ {denotes} the number of paths, with the zeroth path {corresponding to} the line-of-sight (LoS) component and $i\in\{1,\dots,I-1\}$ representing the non-line-of-sight (NLoS) components. Moreover, $\alpha_{i,k}$ denotes the path gain of the $i$-th path and $\mathbf{a}(\hat{\mathbf{y}},\vartheta_{i,k},\varphi_{i,k})$ is the steering vector of the $k$-th user, which depends on the azimuth and elevation angles of departure (AoDs) of the $i$-th path from the final SFIM layer, denoted by $\vartheta_{i,k}$ and $\varphi_{i,k}$, {respectively}.

Taking the top-left meta-atom of the RE of the {final} layer as a reference element, {the $u$-th element of the steering vector can be expressed as}\footnote{The channel model in~\eqref{eq:SV_CM} assumes that the maximum morphing distance (i.e., $\tilde{y}$) is negligible compared to the communication distance between the final SFIM layer and the receivers. As a result, the path gains across all meta-atoms are approximately equal. This assumption is reasonable in practice, as the morphing distance is typically on the order of millimeters, whereas the communication distance is usually on the order of tens of meters or more.}
\begin{equation}
	[\mathbf{a}(\hat{\mathbf{y}},\vartheta,\varphi)]_{u}=\exp\big\{j 2 \pi \big(\psi_{u}(\vartheta,\varphi)+\hat{y}_{u}^{[L]}\sin\vartheta\sin\varphi\big) / \lambda \big\},
\end{equation}
where $\psi_{u}(\vartheta,\varphi)\triangleq c_{x}\operatorname{mod}(u-1,N_{x})\cos\vartheta\sin\varphi+c_{z}\lfloor(u-1)/N_{x}\rfloor\cos\varphi$.

{This channel model can also be adopted for DSIM} by setting $\hat{y}^{[\ell]}_1 = \dots = \hat{y}^{[\ell]}_N \triangleq \hat{y}^{[\ell]}$ for all $\ell \in \mathcal{L}$.

\subsection{Received Signal Model}
The signal received by the $k$-th user can expressed as 
\begin{equation}
    f_{k}=\mathbf{g}_{k}^\mathsf{T}(\hat{\mathbf{y}},\boldsymbol{\phi})\mathbf{x}+n_{k},
\end{equation}
where $n_{k}\sim\mathcal{CN}(0,\sigma_{k}^{2})$ is the additive white Gaussian noise (AWGN).

\section{Problem Formulation\label{sec:prob_form}}

In this section, we formulate the joint optimization problem for the proposed SIM-enabled communication system. Our objective is to maximize the system sum rate by jointly optimizing the transmit beamforming matrix at the BS, the discrete meta-atom responses across all SIM layers, and the continuous morphing parameters of the meta-atoms, subject to practical power, quality-of-service (QoS), quantization, morphing, and inter-layer distance constraints. To this end, we first characterize the signal-to-interference-plus-noise ratio (SINR) and the achievable rate of each user, and then formulate the corresponding sum-rate maximization problems for both SFIM- and DSIM-based architectures.

We can express the SINR of the $k$-th user as
\begin{equation}
	\gamma_{k}(\mathbf{W},\hat{\mathbf{y}},\boldsymbol{\phi})=\frac{J_{k,k}(\hat{\mathbf{y}},\boldsymbol{\phi})}{\sum\nolimits_{i\in\mathcal{K}\setminus\{k\}}J_{k,i}(\hat{\mathbf{y}},\boldsymbol{\phi})+\sigma_{k}^{2}}.
\end{equation}
$J_{k,i}(\hat{\mathbf{y}},\boldsymbol{\phi})\triangleq\vert \mathbf{g}_{k}^\mathsf{T}(\hat{\mathbf{y}},\boldsymbol{\phi}) \mathbf{w}_i \vert^{2}$. Then, the achievable rate of the $k$-th user is defined as
\begin{align}
	& R_{k}(\mathbf{W},\hat{\mathbf{y}},\boldsymbol{\phi}) = \log_{2}\big[1+\gamma_{k}(\mathbf{W},\hat{\mathbf{y}},\boldsymbol{\phi})\big] \notag \\
	= & \log_{2}\Big[\sum\nolimits_{i\in\mathcal{K}} J_{k,i}(\mathbf{W},\hat{\mathbf{y}},\boldsymbol{\phi})+\sigma_{k}^{2}\Big] \notag \\
	& \qquad \qquad {-\log_{2}}\Big[\sum\nolimits_{i\in\mathcal{K}\setminus\{k\}} J_{k,i}(\mathbf{W},\hat{\mathbf{y}},\boldsymbol{\phi})+\sigma_{k}^{2}\Big].
\end{align}
The system sum rate is then given by
\begin{equation}
	R_{\text{sum}}\big(\mathbf{W},\hat{\mathbf{y}},\boldsymbol{\phi}\big)=\sum\nolimits_{k\in\mathcal{K}}R_{k}\big(\mathbf{W},\hat{\mathbf{y}},\boldsymbol{\phi}\big).
\end{equation}

Our main goal is to maximize the sum rate. For the proposed SFIM-based system, this problem can be formulated as follows:
\begin{subequations}
	\label{eq:main_opt}
	\begin{align}
		(\mathcal{P}_1): \ &\underset{\mathbf{W},\hat{\mathbf{y}},\boldsymbol{\Phi}}{\max}\  R_{\text{sum}}\big(\mathbf{W},\hat{\mathbf{y}},\boldsymbol{\phi}\big)\label{eq:obj_fun}\\
		\text{s.t.} \   & \Vert \mathbf{W} \Vert_\mathrm{F}^2 \leq P_{\max},\label{eq:power_cons}\\
		\  & R_{k} (\mathbf{W},\hat{\mathbf{y}},\boldsymbol{\phi}\big) \geq R^\mathrm{th}_k \ \forall k\in\mathcal{K},\label{eq:QoS_cons}\\
		\  & \phi_{n}^{[\ell]}  \in \mathcal{Q} \ \forall n\in\mathcal{N},\ \forall\ell\in\mathcal{L},\label{eq:RIS_cons}\\
		\  & -\tilde{y}\leq\hat{{y}}_{n}^{[\ell]}\leq\tilde{y}\ \forall n\in\mathcal{N},\ \forall\ell\in\mathcal{L},\label{eq:morp_cons} \\
		& d_{n,m}^{[1]}(\hat{y}_{n}^{[1]}) \geq \epsilon \ \forall n\in\mathcal{N},\ \forall m \in\mathcal{M}, \label{eq:dist_cons1} \\
		& d_{n,m}^{[\ell]}(\hat{y}_{m}^{[\ell-1]},\hat{y}_{n}^{[\ell]}) \geq \epsilon \ \forall n,m\in\mathcal{N},\ \forall \ell \in\mathcal{L} \setminus \{ 1\}. \label{eq:dist_cons2} 
	\end{align}
\end{subequations}
In~$(\mathcal{P}_1)$, the constraint~\eqref{eq:power_cons} represents the transmit power constraint, where $P_{\max}$ denotes the maximum power budget. The constraint in~\eqref{eq:QoS_cons} ensures that the achievable rate of the $k$-th user is no smaller than the prescribed threshold $R_{k}^{\mathrm{th}}$, while the constraint~\eqref{eq:RIS_cons} enforces discrete-valued meta-atom responses. The constraint~\eqref{eq:morp_cons} restricts the morphing of each meta-atom to the allowable range. Furthermore, the constraints~\eqref{eq:dist_cons1} and~\eqref{eq:dist_cons2} guarantee that the distances between any transmit antenna and the meta-atoms on the first layer, as well as between meta-atoms on different layers, exceed a minimum threshold $\epsilon$, which is required for the validity of the Rayleigh--Sommerfeld channel model.\footnote{In practice, the inter-layer distances must be sufficiently large to avoid operation in the reactive near-field region.}

For the DSIM-enabled system, the corresponding optimization problem is formulated as
\begin{subequations}
	\label{eq:main_opt2}
	\begin{align}
		(\mathcal{P}_2): \ &\underset{\mathbf{W},\hat{\mathbf{y}},\boldsymbol{\Phi}}{\max}\  R_{\text{sum}}\big(\mathbf{W},\hat{\mathbf{y}},\boldsymbol{\phi}\big)\label{eq:obj_funP2}\\
		\text{s.t.} \   &  \eqref{eq:power_cons}, \eqref{eq:QoS_cons}, \eqref{eq:RIS_cons},\eqref{eq:morp_cons}, \eqref{eq:dist_cons1}, \eqref{eq:dist_cons2}, \\
		& \hat{y}^{[\ell]}_1 = \dots = \hat{y}^{[\ell]}_N \ \forall \ell \in \mathcal{L} \label{eq:DSIM_cons}.
	\end{align}
\end{subequations}

{Solving the problems~$(\mathcal P_1)$ and~$(\mathcal P_2)$ is highly challenging from an optimization-theoretic perspective. The objective function is non-concave due to the multiplicative and highly coupled structure of the cascaded SIM channel, which jointly depends on the transmit beamforming matrix, the discrete meta-atom responses, and the morphing parameters. Moreover, the QoS constraints in~\eqref{eq:QoS_cons} are inherently non-convex, as the SINR expressions involve fractional quadratic forms, while the discrete constraint~\eqref{eq:RIS_cons} renders the feasible set non-convex and non-smooth. These challenges are further exacerbated by the strong coupling of the optimization variables across both the objective function and the constraints, which precludes the direct application of standard convex optimization techniques.}

\vspace{-0.3cm}
\section{Proposed Solution\label{sec:prop_sol} }
In this section, we develop an alternating optimization framework that integrates gradient projection, penalty-based optimization, and successive convex approximation techniques to compute stationary solutions to~$(\mathcal P_1)$ and~$(\mathcal P_2)$.
 
 \vspace{-0.3cm}

\subsection{Morphing Optimization}
We begin by presenting the morphing optimization for~$(\mathcal P_1)$, and then outline the necessary modifications for the morphing optimization for the problem in~$(\mathcal P_2)$.

In the following, we assume that the values of $\mathbf{W}$ and $\boldsymbol{\phi}$ are fixed, and we aim to update $\hat{\mathbf{y}}$. This optimization sub-problem can be written as
\begin{subequations}
\label{eq:y_opt}
\begin{align}
\underset{\hat{\mathbf{y}}}{\max}\  & R_{\text{sum}}\big(\hat{\mathbf{y}}\big)\label{eq:y_obj_fun}\\
\text{s.t.} \   & R_{k} (\hat{\mathbf{y}}) \geq R^\mathrm{th}_k \ \forall k\in\mathcal{K},\label{eq:y_QoS_cons}\\
 & \eqref{eq:morp_cons}, \eqref{eq:dist_cons1}, \eqref{eq:dist_cons2}.
\end{align}
\end{subequations}

To handle the challenging constraint~\eqref{eq:y_QoS_cons}, we introduce the following penalty function:
\begin{equation}
    R_\text{pen} (\hat{\mathbf{y}}) \triangleq \sum\nolimits_{k \in \mathcal{K}} \big(R^\mathrm{th}_k -  R_{k} (\hat{\mathbf{y}}) + \mu_k \big)^2,
\end{equation}
where $\mu_k \geq 0$ is a slack variable introduced to remove the penalty terms associated with the satisfied constraints for all $k \in \mathcal{K}$. We then introduce the following augmented function:
\begin{equation}
    R_\text{aug} (\hat{\mathbf{y}}) \triangleq R_\text{sum}(\hat{\mathbf{y}})  - \frac{1}{2 \omega}R_\text{pen} (\hat{\mathbf{y}}) ,
\end{equation}
where $ \omega > 0$ denotes a tuning parameter that regulates the strength of the penalty term. 

Then, we can find the gradient $\nabla_{\hat{\mathbf{y}}} R_\text{aug} (\hat{\mathbf{y}})$ as follows:
\begin{equation}
    \nabla_{\hat{\mathbf{y}}} R_\text{aug} (\hat{\mathbf{y}}) = \sum_{k \in \mathcal{K}} \Big[1+ \frac{1}{\omega} (R^\mathrm{th}_k -  R_{k} (\hat{\mathbf{y}}) + \mu_k) \Big] \nabla_{\hat{\mathbf{y}}} R_k (\hat{\mathbf{y}}).
\end{equation}
We can obtain the gradient of $R_{k}(\hat{\mathbf{y}})$ with respect to $\hat{\mathbf{y}}$ as 
\begin{multline}
\nabla_{\hat{\mathbf{y}}}R_{k}(\hat{\mathbf{y}})=\frac{1}{\ln(2)}\Bigg[\frac{\sum\nolimits_{i \in \mathcal{K}} \nabla_{\hat{\mathbf{y}}} J_{k,i} (\hat{\mathbf{y}})\!}{\sum\nolimits_{i \in \mathcal{K}} J_{k,i} (\hat{\mathbf{y}})\!+  \!\sigma_{k}^{2}} \\
- \frac{\sum\nolimits_{i \in \mathcal{K} \setminus \{k \}} \nabla_{\hat{\mathbf{y}}} J_{k,i} (\hat{\mathbf{y}})\!}{\sum\nolimits_{i \in \mathcal{K} \setminus \{k \}} J_{k,i} (\hat{\mathbf{y}})\!+  \!\sigma_{k}^{2}}\Big].
\end{multline}
Next, we find the gradient $\nabla_{\hat{\mathbf{y}}}J_{k,i}(\hat{\mathbf{y}})=\Big[\frac{\partial J_{k,i}(\hat{\mathbf{y}})}{\partial\hat{y}_{1}^{[1]}},\dots,\frac{\partial J_{k,i}(\hat{\mathbf{y}})}{\partial\hat{y}_{N}^{[1]}},\dots,\frac{\partial J_{k,i}(\hat{\mathbf{y}})}{\partial\hat{y}_{1}^{[L]}},\dots,\frac{\partial J_{k,i}(\hat{\mathbf{y}})}{\partial\hat{y}_{N}^{[L]}}\Big]^\mathsf{T}$.

\paragraph{For\textit{\emph{ $\ell\in\mathcal{L}\setminus\{L\}$}}}

\begin{multline}
\tfrac{\partial J_{k,i}(\hat{\mathbf{y}})}{\partial\hat{y}_{n}^{[\ell]}}=2\Re\bigg\{\bigg[\big(\mathbf{c}_{k}^{[\ell]}(\hat{\mathbf{y}})\big)^\mathsf{T}\bigg\{\frac{\partial\boldsymbol{\Omega}^{[\ell+1]}(\hat{\mathbf{y}})}{\partial\hat{y}_{n}^{[\ell]}}\boldsymbol{\Phi}^{[\ell]}\boldsymbol{\Omega}^{[\ell]}(\hat{\mathbf{y}})\\
+\boldsymbol{\Omega}^{[\ell+1]}(\hat{\mathbf{y}})\boldsymbol{\Phi}^{[\ell]}\frac{\partial\boldsymbol{\Omega}^{[\ell]}(\hat{\mathbf{y}})}{\partial\hat{y}_{n}^{[\ell]}}\bigg\}\mathbf{e}_i^{[\ell]}(\hat{\mathbf{y}})\bigg]^{*}\mathbf{g}^\mathsf{T}_{k}(\hat{\mathbf{y}})\mathbf{w}_i\bigg\},\label{eq:dGk}
\end{multline}
where $(\mathbf{c}_{k}^{[\ell]}(\hat{\mathbf{y}}))^\mathsf{T}\triangleq(\mathbf{h}_{k}(\hat{\mathbf{y}}))^\mathsf{T}\big(\prod_{u=L}^{\ell+2}\boldsymbol{\Phi}^{(u)}\boldsymbol{\Omega}^{(u)}(\hat{\mathbf{y}})\big)\boldsymbol{\Phi}^{[\ell+1]}$ and $\mathbf{e}_i^{[\ell]}(\hat{\mathbf{y}})\triangleq (\prod_{u=\ell-1}^{[1]}\boldsymbol{\Phi}^{(u)}\boldsymbol{\Omega}^{(u)}(\hat{\mathbf{y}})) \mathbf{w}_i$.

We can observe from~\eqref{eq:dGk} that only the $n$-th column of $\partial\boldsymbol{\Omega}^{[\ell]}(\hat{\mathbf{y}})/\partial\hat{y}_{n}^{[\ell]}$ can have nonzero elements. Therefore, the product $\boldsymbol{\Omega}^{[\ell+1]}(\hat{\mathbf{y}})\boldsymbol{\Phi}^{[\ell]}\partial\boldsymbol{\Omega}^{[\ell]}(\hat{\mathbf{y}})/\partial\hat{y}_{n}^{[\ell]}$ can be expressed as a matrix whose elements are all zero except for the $n$-th column, which is $\phi_{n}^{[\ell]}\boldsymbol{\Omega}^{[\ell+1]}(\hat{\mathbf{y}})\mathbf{z}_{n}^{[\ell]}(\hat{\mathbf{y}})$, where 
\begin{multline}
[\mathbf{z}_{n}^{[\ell]}({\hat{\mathbf{y}}})]_{m}\triangleq\frac{\partial p_{n,m}^{[\ell]}({\hat{\mathbf{y}}})}{\partial\hat{y}_{n}^{[\ell]}}q_{n,m}^{[\ell]}({\hat{\mathbf{y}}})r_{n,m}^{[\ell]}({\hat{\mathbf{y}}})+p_{n,m}^{[\ell]}({\hat{\mathbf{y}}})\\
\times\frac{\partial q_{n,m}^{[\ell]}({\hat{\mathbf{y}}})}{\partial\hat{y}_{n}^{[\ell]}}r_{n,m}^{[\ell]}({\hat{\mathbf{y}}})+p_{n,m}^{[\ell]}({\hat{\mathbf{y}}})q_{n,m}^{[\ell]}({\hat{\mathbf{y}}})\frac{\partial r_{n,m}^{[\ell]}({\hat{\mathbf{y}}})}{\partial\hat{y}_{n}^{[\ell]}},\label{eq:beg}
\end{multline}
with 
\begin{subequations}
	\begin{align}
		& \frac{\partial p_{n,m}^{[\ell]}(\hat{\mathbf{y}})}{\partial\hat{y}_{n}^{[\ell]}} = \frac{\tilde{A}^{[\ell]}\big( {\rho_{n,m}^{[\ell]}}-(\hat{y}_{n}^{[\ell]}+\xi_{m}^{[\ell]})^{2}\big)}{\big( {\rho_{n,m}^{[\ell]}}+(\hat{y}_{n}^{[\ell]}+\xi_{m}^{[\ell]})^{2}\big)^{2}}, \\
		& \frac{\partial q_{n,m}^{[\ell]}(\hat{\mathbf{y}})}{\partial\hat{y}_{n}^{[\ell]}} = -\frac{\hat{y}_{n}^{[\ell]}+\xi_{m}^{[\ell]}}{2\pi\big({\rho_{n,m}^{[\ell]}}+(\hat{y}_{n}^{[\ell]}+\xi_{m}^{[\ell]})^{2}\big)^{3/2}}, \\
		& \frac{\partial r_{n,m}^{[\ell]}(\hat{\mathbf{y}})}{\partial\hat{y}_{n}^{[\ell]}} = j \frac{2\pi(\hat{y}_{n}^{[\ell]}+\xi_{m}^{[\ell]})}{\lambda\big({\rho_{n,m}^{[\ell]}}+(\hat{y}_{n}^{[\ell]}+\xi_{m}^{[\ell]})^{2}\big)^{1/2}} \notag \\
		& \qquad \qquad \times \exp\Big(j\frac{2\pi \big({\rho_{n,m}^{[\ell]}}+(\hat{y}_{n}^{[\ell]}+\xi_{m}^{[\ell]})^{2}\big)^{1/2}}{\lambda}\Big).
	\end{align}
\end{subequations}
Moreover, $\rho_{n,m}^{[\ell]}\triangleq(x_{n}^{[\ell]}-x_{m}^{[\ell-1]})^{2}+(z_{n}^{[\ell]}-z_{m}^{[\ell-1]})^{2}$, and
\begin{equation}
\xi_{m}^{[\ell]}\triangleq\begin{cases}
\bar{y}^{[1]}, & \ell=1,\\
\bar{y}^{[\ell]}-\hat{y}_{m}^{[\ell-1]}, & \ell\in \mathcal{L} \setminus \{1\}.
\end{cases}\label{eq:end}
\end{equation}

Similarly, only the $n$-th row of $\partial\boldsymbol{\Omega}^{[\ell+1]}(\hat{\mathbf{y}}) / \partial\hat{y}_{n}^{[\ell]}$ can have nonzero elements. Therefore, the product $\frac{\partial\boldsymbol{\Omega}^{[\ell+1]}(\hat{\mathbf{y}})}{\partial\hat{y}_{n}^{[\ell]}}\boldsymbol{\Phi}^{[\ell]}\boldsymbol{\Omega}^{[\ell]}(\hat{\mathbf{y}})$ can be expressed as a matrix whose elements are all zero except for the $n$-th row, which is $\phi_{n}^{[\ell]}\big(\boldsymbol{\eta}_{n}^{[\ell]}(\hat{\mathbf{y}})\big)^\mathsf{T}\boldsymbol{\Omega}^{[\ell]}(\hat{\mathbf{y}})$, where 
\begin{align}
[\boldsymbol{\eta}_{n}^{[\ell]}(\hat{\mathbf{y}})]_{m} \triangleq & \frac{\partial p_{m,n}^{[\ell+1]}(\hat{\mathbf{y}})}{\partial\hat{y}_{n}^{[\ell]}}q_{m,n}^{[\ell+1]}(\hat{\mathbf{y}})r_{m,n}^{[\ell+1]}(\hat{\mathbf{y}})\notag \\
& + p_{m,n}^{[\ell+1]}(\hat{\mathbf{y}}) \frac{\partial q_{m,n}^{[\ell+1]}(\hat{y}_{n}^{[\ell]})}{\partial\hat{y}_{n}^{[\ell]}}r_{m,n}^{[\ell+1]}(\hat{y}_{n}^{[\ell]}) \notag \\
& + p_{m,n}^{[\ell+1]}(\hat{y}_{n}^{[\ell]})q_{m,n}^{[\ell+1]}(\hat{y}_{n}^{[\ell]})\frac{\partial r_{m,n}^{[\ell+1]}(\hat{y}_{n}^{[\ell]})}{\partial\hat{y}_{n}^{[\ell]}},
\end{align}
with 
\begin{subequations}
	\begin{align}
		& \frac{\partial p_{m,n}^{[\ell+1]}(\hat{y}_{n}^{[\ell]})}{\partial\hat{y}_{n}^{[\ell]}} = \frac{\tilde{A}^{[\ell]}\big((\hat{y}_{n}^{[\ell]}-\varsigma_{m}^{[\ell]})^{2}-\rho_{m,n}\big)}{\big({\rho_{m,n}^{[\ell]}}+(\hat{y}_{n}^{[\ell]}-\varsigma_{m}^{[\ell]})^{2}\big)^{2}}, \\
		& \frac{\partial q_{m,n}^{[\ell+1]}(\hat{\mathbf{y}})}{\partial\hat{y}_{n}^{[\ell]}} = -\frac{\hat{y}_{n}^{[\ell]}-\varsigma_{m}^{[\ell]}}{2\pi\big( {\rho_{m,n}^{[\ell]}} +(\hat{y}_{n}^{[\ell]}-\varsigma_{m}^{[\ell]})^{2}\big)^{3/2}}, \\
		& \frac{\partial r_{m,n}^{[\ell+1]}(\hat{\mathbf{y}})}{\partial\hat{y}_{n}^{[\ell]}} = j\frac{2\pi(\hat{y}_{n}^{[\ell]}-\varsigma_{m}^{[\ell]})}{\lambda \big( {\rho_{m,n}^{[\ell]}} +(\hat{y}_{n}^{[\ell]}-\varsigma_{m}^{[\ell]})^{2}\big)^{1/2}} \notag \\
		& \qquad \qquad \times \exp\big\{j2\pi \big( {\rho_{m,n}^{[\ell]}} +(\hat{y}_{n}^{[\ell]}-\varsigma_{m}^{[\ell]})^{2}\big)^{1/2}\big/\lambda\big\},\!
	\end{align}
\end{subequations}
and $\varsigma_{m}^{[\ell]}\triangleq\bar{y}^{[\ell+1]}+\hat{y}_{m}^{[\ell+1]}.$

\paragraph{For $\ell=L$}
\begin{multline}
\frac{\partial J_{k,i}(\hat{\mathbf{y}})}{\partial {\hat{y}_{n}^{[\ell]}}}=2\Re \bigg\{ \bigg[ \bigg\{ \bigg(\frac{\partial\mathbf{h}_{k}(\hat{\mathbf{y}})}{\partial\hat{y}_{n}^{[L]}} \bigg)^\mathsf{T}\boldsymbol{\Phi}^{[L]}\boldsymbol{\Omega}^{[L]}(\hat{\mathbf{y}})\\
+\mathbf{h}_{k}^\mathsf{T}(\hat{\mathbf{y}})\boldsymbol{\Phi}^{[L]}\frac{\partial\boldsymbol{\Omega}^{[L]}(\hat{\mathbf{y}})}{\partial\hat{y}_{n}^{[L]}}\bigg\} {\mathbf{e}_i^{[L]}}(\hat{\mathbf{y}})\bigg]^{*}\mathbf{g}^\mathsf{T}_{k}(\hat{\mathbf{y}}){\mathbf{w}}_i\bigg\},
\end{multline}
We can obtain $\partial\boldsymbol{\Omega}^{[L]}(\hat{\mathbf{y}}) / \partial\hat{y}_{n}^{[L]}$ using~\eqref{eq:beg} and~\eqref{eq:end}. On the other hand, we can obtain $\partial\mathbf{h}_{k}(\hat{\mathbf{y}}) / \partial\hat{y}_{n}^{[L]}$ as 
\begin{multline}
\frac{\partial\mathbf{h}_{k}(\hat{\mathbf{y}})}{\partial {\hat{y}_{n}^{[\ell]}}} = j \frac{2\pi}{\lambda}\sum\nolimits_{i=0}^{I-1}\alpha_{i,k}\sin\vartheta_{i,k}\sin\varphi_{i,k}\exp\bigg(j\frac{2\pi}{\lambda}\\
\times\big\{\psi_{n}(\vartheta_{i,k},\varphi_{i,k})+\hat{y}_{n}^{[L]}\sin\vartheta_{i,k}\sin\varphi_{i,k}\big\}\bigg),
\end{multline}

Thus, we can update $\hat{\mathbf{y}}$ as follows: 
\begin{equation}
\hat{\mathbf{y}}^{{(t)}}\leftarrow\hat{\mathbf{y}}^{(t-1)}+\delta_{\hat{\mathbf{y}}}\nabla_{\hat{\mathbf{y}}}R_{\text{aug}} \big(\hat{\mathbf{y}}^{(t-1)} \big),\label{eq:y_update}
\end{equation}
where $\hat{\mathbf{y}}^{{(t)}}$ is the value of $\hat{\mathbf{y}}$ at the $t$-th iteration. Also, $\delta_{\hat{\mathbf{y}}}$ is the step size. To select an appropriate step size, we start with an initial step size, then keep decreasing it until the following condition is satisfied~\cite{2025_Bahingayi}:
\begin{equation}
    R_{\text{aug}}(\hat{\mathbf{y}}^{{(t)}}) \geq R_{\text{aug}} \big(\hat{\mathbf{y}}^{(t-1)} \big) + \varepsilon \Vert \hat{\mathbf{y}}^{{(t)}} - \hat{\mathbf{y}}^{(t-1)} \Vert_2^2,
\end{equation}
where $\varepsilon \geq 0$ is a small parameter.

Next, we project $\hat{\mathbf{y}}^{{(t)}}$ onto the feasible set. The constraint set~\eqref{eq:morp_cons} can be written as
\begin{equation}
    - \tilde{y} \mathbf{1}_{NL \times 1} \leq \hat{\mathbf{y}} \leq \tilde{y} \mathbf{1}_{NL \times 1}.
    \label{eq:new_morp_cons}
\end{equation}
We then re-express the constraint set~\eqref{eq:dist_cons1} as
\begin{equation}
       \hat{y}_n^{[1]} \geq \big\{\max \big(0,\epsilon^2- \min_m (\rho^{[1]}_{n,m})\big) \big\}^{1/2} - \bar{y}^{[1]} \triangleq \zeta_n^{[1]} \ \forall n \in \mathcal{N}.
       \label{eq:y2}
\end{equation}
Since the surfaces are assumed to be perfectly aligned along the $x$-$z$ plane, we can re-express the constraint set~\eqref{eq:dist_cons2} as
\begin{equation}
        {\hat{y}_n^{[\ell]} - \hat{y}_n^{[\ell-1]}  \geq  \big\{\max ( 0, \epsilon^2 - \rho^{[\ell]}_{n,n}) \big\}^{1/2}  - \bar{y}^{[\ell]}  \triangleq \zeta_n^{[\ell]},}
\end{equation}
which implies that
\begin{equation}
          (\boldsymbol{\chi}_{n}^{[\ell]} )^\mathsf{T} \hat{\mathbf{y}}   \geq  \zeta_n^{[\ell]} \ \forall n \in \mathcal{N}, \  \forall \  \ell \in \mathcal{L} \setminus \{ 1\},
        \label{eq:y1}
\end{equation}
where $\boldsymbol{\chi}_{n}^{[\ell]}$ is an $NL \times 1$ vector all entries of which are equal to zero except the $\big((\ell -1)N+n\big)$-th and the $\big((\ell -2)N+n\big)$-th entries, which are set to $1$ and $-1$, respectively. Then, by defining 
\begin{equation}
    \boldsymbol{\Delta} \triangleq \begin{bmatrix}
    \begin{bmatrix}
        \mathbf{I}_{N} & \mathbf{0}_{N} & \dots & \mathbf{0}_{N}
    \end{bmatrix}\\
        (\boldsymbol{\chi}_{1}^{[2]} )^\mathsf{T} \\
        \vdots \\
        {(\boldsymbol{\chi}_{N}^{[2]} )^\mathsf{T}} \\
        \vdots \\
        (\boldsymbol{\chi}_{1}^{[L]} )^\mathsf{T} \\
        \vdots \\
        (\boldsymbol{\chi}_{N}^{[L]} )^\mathsf{T} \\
    \end{bmatrix}, \ \    \boldsymbol{\zeta} \triangleq \begin{bmatrix}
        \zeta_{1}^{[1]}  \\
        \vdots \\
       \zeta_{N}^{[1]} \\
        \vdots \\
        \zeta_{1}^{[L]} \\
        \vdots \\
        \zeta_{N}^{[L]}  \\
    \end{bmatrix},
\end{equation}
we can combine the constraint sets~\eqref{eq:dist_cons1} and~\eqref{eq:dist_cons2} to form a single vector constraint, i.e.,
\begin{equation}
    \boldsymbol{\zeta} \leq \boldsymbol{\Delta} \hat{\mathbf{y}}.
    \label{eq:new_dist_cons}
\end{equation}

Therefore, the update of $\hat{\mathbf{y}}^{{(t)}}$ obtained from~\eqref{eq:y_update} can be projected onto the feasible set by solving the following convex optimization problem:
\begin{subequations}
\label{eq:y_proj}
\begin{align}
\underset{\hat{\mathbf{y}}}{{\min}}\  & \Vert \hat{\mathbf{y}} - \hat{\mathbf{y}}^{{(t)}}\Vert_2^2\\
\text{s.t.} \   & \eqref{eq:new_morp_cons}, \eqref{eq:new_dist_cons},
\end{align}
\end{subequations}
which can be solved using the CVX toolbox~\cite{CVX}.

The morphing update of~\eqref{eq:main_opt2} entails an additional projection onto the constraint set~\eqref{eq:DSIM_cons}. This projection can be simply performed as follows~\cite{2006_strang}:
\begin{equation}
    \hat{y}^{[\ell]}_n \leftarrow \frac{1}{N} \sum\nolimits_{i \in \mathcal{N}} \hat{y}^{[\ell]}_i \ \forall n \in \mathcal{N} \ \forall \ell \in \mathcal{L}.
    \label{eq:add_projy}
\end{equation}

After projecting onto the feasible set, we update the slack variables $\{ \mu_k\}$ as follows~\cite{2025_Magbool2}:
\begin{equation}
    \mu_k \leftarrow \max \{ 0, { R_k (\hat{\mathbf{y}}) - R^{\text{th}}_k}  \} \ \forall k \in \mathcal K.
    \label{eq:mu_update}
\end{equation}
For a given $\hat{\mathbf{y}}$, the update in \eqref{eq:mu_update} is the closed-form solution of
$\min_{\mu_k\ge 0} (R_k^{\mathrm{th}}-R_k(\hat{\mathbf{y}})+\mu_k)^2$, and yields zero penalty whenever $R_k(\hat{\mathbf y}) \geq R_k^{\mathrm{th}}$. We then define the QoS violation measure $\wp_k \triangleq \max\{0, R^{\text{th}}_k - R_k(\hat{\mathbf{y}})\}$. If $\max_{k\in\mathcal{K}} \wp_k > 0$, we increase the penalty by decreasing $\omega$ according to
\begin{equation}
    \omega \leftarrow \kappa \omega ,
\end{equation}
where $0 < \kappa < 1$. \textbf{Algorithm~\ref{alg:y_update}} summarizes the morphing update.

\begin{algorithm}[t]
\caption{Morphing update.}
\label{alg:y_update}

\KwIn{ $\mathbf{W}$, $\boldsymbol{\phi}$, $\hat{\mathbf{y}}^{(t-1)}$, $\omega$, $\kappa$ }

Update $\hat{\mathbf{y}}^{{(t)}}$ using~\eqref{eq:y_update}\;

Project onto the feasible set by solving~\eqref{eq:y_proj}\;

If a solution satisfying~\eqref{eq:main_opt2} is required, project onto the additional constraint set~\eqref{eq:DSIM_cons} using~\eqref{eq:add_projy}\;

Update $\mu_k \ \forall k \in \mathcal{K}$ using~\eqref{eq:mu_update}\;

{Define $\wp_k \triangleq \max\{0, R^{\text{th}}_k - R_k(\hat{\mathbf{y}})\} \ \forall k \in \mathcal K$}\;
\While{ {$\max_{k\in\mathcal{K}} \wp_k > 0$} }{

Update $\omega \leftarrow \kappa \omega$\; 

Repeat steps 1-{5}\;

}

\KwOut{  $\hat{\mathbf{y}}^{{(t)}}$ }
\end{algorithm}

\vspace{-0.5cm}

\subsection{Transmit Beamformer Optimization}
{Fixing $\hat{\mathbf{y}}$ and $\boldsymbol{\phi}$, the transmit beamforming design reduces to a non-convex sum-rate maximization problem with respect to the precoding matrix $\mathbf{W}$. The resulting subproblem is formulated as}
\begin{subequations}\label{eq:p_opt}
	\begin{align}
		\max_{\mathbf{W}} \ \ & R_\text{sum}(\mathbf{W}) \label{obj_fun_p}\\
		\text{s.t.}\ \ & \eqref{eq:power_cons},\ \ R_k(\mathbf{W}) \ge R_k^\text{th}\ \forall k\in\mathcal K. \label{cons2_p}
	\end{align}
\end{subequations}

We introduce auxiliary slack variables $\{\tau_{i,k}\}$ and $\{\varpi_{i,k}\}$ satisfying
\begin{equation}\label{eq:sv1}
	\tau_{i,k}\le |\mathbf g_k^\mathsf{T}\mathbf w_i|^2\ \forall i,k\in\mathcal K,
\end{equation}
and
\begin{equation}\label{eq:sv2}
	|\mathbf g_k^\mathsf{T}\mathbf w_i|^2\le \varpi_{i,k}\ \forall i,k\in\mathcal K.
\end{equation}
{Note that~\eqref{eq:sv2} represents the epigraph of a convex quadratic function and is therefore convex.}

Using \eqref{eq:sv1}–\eqref{eq:sv2}, the {sum-rate objective in~\eqref{obj_fun_p} admits the following global lower bound:}
\begin{align}\label{eq:W_opt1}
	R_\text{sum}(\mathbf{W}) \ge & \sum \nolimits_{k\in\mathcal K}\Big[\log_2\Big(\sum \nolimits_{i\in\mathcal K}\tau_{i,k}+\sigma_k^2\Big) \notag \\
	& \qquad - \log_2\Big(\sum \nolimits_{i\in\mathcal K\setminus\{k\}}\varpi_{i,k}+\sigma_k^2\Big)\Big]. 
\end{align}
{The second logarithmic term on the right-hand side of~\eqref{eq:W_opt1} is concave in $\{\varpi_{i,k}\}$ and is handled via SCA. Let $\mathscr S_k^{(t-1)}\triangleq\sum_{i\in\mathcal K\setminus\{k\}}\varpi_{i,k}^{(t-1)}+\sigma_k^2$. A first-order approximation at iteration $t$ yields
\begin{align}
	& \tilde R_\text{sum}^{(t)}(\{\tau_{i,k}\},\{\varpi_{i,k}\})=\sum \nolimits_{k\in\mathcal K}\Big[\log_2\Big(\sum \nolimits_{i\in\mathcal K}\tau_{i,k}+\sigma_k^2\Big) \notag \\
	& -\log_2\big(\mathscr S_k^{(t-1)} \big)-\frac{1}{\ln 2}\frac{\sum_{i\in\mathcal K\setminus\{k\}}(\varpi_{i,k}-\varpi_{i,k}^{(t-1)})}{\mathscr S_k^{(t-1)}}\Big].
\end{align}
Moreover, since $|\mathbf g_k^\mathsf{T}\mathbf w_i|^2$ is convex in $\mathbf w_i$, a global affine underestimator of \eqref{eq:sv1} at $\mathbf w_i^{(t-1)}$ is given by}
\begin{equation}\label{eq:tr_sv1}
	\tau_{i,k}\le 2\Re\{\mathbf g_k^\mathsf{T}\mathbf w_i^{(t-1)}\mathbf w_i^\mathsf{H}\mathbf g_k^*\}-|\mathbf g_k^\mathsf{T}\mathbf w_i^{(t-1)}|^2\ \forall i,k\in\mathcal K.
\end{equation}

{Using \eqref{eq:sv1}–\eqref{eq:sv2}, a conservative lower bound on the $k$-th user rate is}
\begin{equation}
	R_k(\mathbf W)\ge \log_2\Big(1+\frac{\tau_{k,k}}{\sum_{i\in\mathcal K\setminus\{k\}}\varpi_{i,k}+\sigma_k^2}\Big),
\end{equation}
{which leads to the sufficient QoS constraint}
\begin{equation}\label{eq:new_cons_W}
	\tau_{k,k}\ge  \big(2^{R_k^\text{th}}-1\big)\Big(\sum \nolimits_{i\in\mathcal K\setminus\{k\}}\varpi_{i,k}+\sigma_k^2\Big)\ \forall k\in\mathcal K.
\end{equation}

Consequently, the convex SCA subproblem solved at iteration $t$ is
\begin{subequations}\label{eq:W_FO}
	\begin{align}
		\max_{\mathbf W,\{\tau_{i,k}\},\{\varpi_{i,k}\}}\ \ & \tilde R_\text{sum}^{(t)}(\{\tau_{i,k}\},\{\varpi_{i,k}\})\\
		\text{s.t.}\ \ & \eqref{eq:power_cons},\ \eqref{eq:sv2},\ \eqref{eq:tr_sv1},\ \eqref{eq:new_cons_W},
	\end{align}
\end{subequations}
{which constitutes a disciplined convex program and can be efficiently solved using standard solvers such as CVX.}

\vspace{-0.3cm}

\subsection{Meta-Atom Response Optimization}
Next, we optimize the meta-atom responses for the $\ell'$-th layer. In this process, we assume that the values of $\mathbf{W}$, $\hat{\mathbf{y}}$, and $\boldsymbol{\Phi}^{[\ell]}$ for all $\ell \in \mathcal{L} \setminus \{\ell'\}$ are fixed, and we aim to update $\boldsymbol{\Phi}^{[\ell']}$. The corresponding optimization sub-problem is
\begin{subequations}
	\begin{align}
		\max_{  \boldsymbol{\Phi}^{[\ell']} }    \ \      & R_\text{sum} (  \boldsymbol{\Phi}^{[\ell']}) \label{obj_fun_phi} \\
		\text{s.t.} \ \ & R_k (\boldsymbol{\Phi}^{[\ell']}) \geq R^\text{th}_k, \ \forall k \in \mathcal{K}, \label{eq:phi_qos_con} \\
		& \boldsymbol{\phi}^{[\ell']}_n \in \mathcal{Q} \ \forall n \in \mathcal{N}. \label{eq:quan_l} 
	\end{align} 
	\label{eq:phi_opt}
\end{subequations}%
We start by expressing the $k$-th user’s rate as
\begin{align}
	& R_k ( \boldsymbol{\Phi}^{[\ell']} )  = \log_2 \Big(\sum\nolimits_{{i} \in \mathcal{K}} \big|(\tilde{\mathbf{g}}_k^{[\ell']})^\mathsf{T}  \boldsymbol{\Phi}^{[\ell']} \tilde{\mathbf{w}}^{[\ell']}_{{i}}\big|^2 + \sigma_k^2 \Big) \notag \\
	& \quad -\log_2 \Big(\sum\nolimits_{{i} \in \mathcal{K} \setminus \{ k\}}  \big|(\tilde{\mathbf{g}}_k^{[\ell']})^\mathsf{T}  \boldsymbol{\Phi}^{[\ell']} \tilde{\mathbf{w}}^{[\ell']}_{{i}}\big|^2 + \sigma_k^2 \Big),
\end{align}
where $(\tilde{\mathbf{g}}_k^{[\ell']})^\mathsf{T} \triangleq \mathbf{h}_k^\mathsf{T} \prod_{\ell = L}^{\ell'+1} \boldsymbol{\Phi}^{[\ell]} \boldsymbol{\Omega}^{[\ell]}$ and $\tilde{\mathbf{w}}_i^{[\ell']} \triangleq \boldsymbol{\Omega}^{[\ell']} \!\left({\prod_{\ell = \ell'-1}^{1}} \boldsymbol{\Phi}^{[\ell]} \boldsymbol{\Omega}^{[\ell]}\right) \mathbf{w}_i$. Then, we introduce the slack variables $\{ \nu_{i,k} \}$ and $\{ \varrho_{i,k} \}$ such that
\begin{equation}
	\nu_{i,k} \leq \big|(\tilde{\mathbf{g}}_k^{[\ell']})^\mathsf{T}  \boldsymbol{\Phi}^{[\ell']} \tilde{\mathbf{w}}^{[\ell']}_i\big|^2 ,
	\label{eq:NINC}
\end{equation}
and 
\begin{equation}
	\varrho_{i,k} \geq \big|(\tilde{\mathbf{g}}_k^{[\ell']})^\mathsf{T}  \boldsymbol{\Phi}^{[\ell']} \tilde{\mathbf{w}}^{[\ell']}_i\big|^2.
	\label{conp1}
\end{equation}
A lower bound on the objective function~\eqref{obj_fun_phi} can be obtained as
\begin{align}
	R_\text{sum} (  \boldsymbol{\Phi}^{[\ell']})  & \geq \sum\nolimits_{k \in \mathcal{K}} \Big[ \log_2 \Big(\sum\nolimits_{i \in \mathcal{K}} \nu_{i,k} + \sigma_k^2 \Big) \notag \\
	& \quad - \log_2 \Big(\sum\nolimits_{i \in \mathcal{K} \setminus \{ k \} } \varrho_{i,k}  + \sigma_k^2 \Big) \Big].
\end{align}
Using the SCA approach, the objective function at the $t$-th iteration is given by the first-order Taylor expansion
\begin{align}
	& \bar{R}^{{(t)}}_\text{sum} ( \{ \nu_{i,k} \}, \{ \varrho_{i,k} \})   \triangleq \sum\nolimits_{k \in \mathcal{K}} \Bigg[ \log_2 \Big(\sum\nolimits_{i \in \mathcal{K}} \nu_{i,k} + \sigma_k^2 \Big) \notag \\
	& \quad { - \log_2 \Big(\sum\nolimits_{i \in \mathcal{K} \setminus \{ k \}} \varrho_{i,k}^{(t-1)} + \sigma_k^2\Big)} \notag \\
	& \quad \qquad \qquad \qquad  - \frac{1}{\ln 2} \frac{{\sum\nolimits_{i \in \mathcal{K} \setminus \{ k\}} (\varrho_{i,k} - \varrho_{i,k}^{(t-1)})}}{{\sum\nolimits_{j \in \mathcal{K} \setminus \{ k \}} \varrho_{j,k}^{(t-1)} + \sigma_k^2}} \Bigg].
\end{align}
The non-convex constraint~\eqref{eq:NINC} is handled via SCA as
\begin{multline}
	\nu_{i,k} \leq 2 \Re \big\{ \big(\tilde{\mathbf{g}}_k^{{[\ell']}} \big)^\mathsf{T}  \boldsymbol{\Phi}^{{[\ell'](t-1)}} \tilde{\mathbf{w}}^{{[\ell']}}_i \big(\tilde{\mathbf{w}}^{{[\ell']}}_i\big)^\mathsf{H} \big(\boldsymbol{\Phi}^{{[\ell'](t-1)}}\big)^*  \\
	\times \big(\tilde{\mathbf{g}}_k^{{[\ell']}} \big)^* \big\} - \big|(\tilde{\mathbf{g}}_k^{{[\ell']}})^\mathsf{T}  \boldsymbol{\Phi}^{{[\ell'](t-1)}} \tilde{\mathbf{w}}^{{[\ell']}}_i\big|^2,
	\label{conp2}
\end{multline}
which constitutes a global lower bound that is tight at $\boldsymbol{\Phi}^{[\ell'](t-1)}$.
Next, we focus {on} the constraint set~\eqref{eq:phi_qos_con}, which can be conservatively lower-bounded as
\begin{equation}
	R_k (\boldsymbol{\Phi}^{[\ell']}) \geq \log_2 \bigg( 1 + \frac{\nu_{k,k}}{{\sum\nolimits_{i \in \mathcal{K} \setminus \{ k\}} \varrho_{i,k}} + \sigma_k^2 } \bigg),
\end{equation}
and equivalently rewritten as
\begin{equation}
	\nu_{k,k} \geq \big(2^{R_k^\text{th}} -1 \big) \Big( \sum\nolimits_{i \in \mathcal{K} \setminus \{ k\}} \varrho_{i,k} + \sigma_k^2 \Big),
	\label{conp3}
\end{equation}
which is linear in the optimization variables.

Finally, a widely used method in the literature to handle the quantization constraint~\eqref{eq:quan_l} is to first solve the continuous meta-atom response problem and then map the obtained phases to their nearest neighbors in the quantized set. However, this approach suffers from two main issues. First, convergence to a stationary point is not guaranteed. Second, the rate constraints of the quantized problem may be violated due to the mismatch introduced by the nearest-neighbor projection. To address these issues, we propose a novel approach in this work to handle the quantization constraint. For that, we introduce the vector $\mathbf{q} \in \mathbb{C}^{U\times 1}$ containing all elements of $\mathcal{Q}$, where $U$ denotes its cardinality. The constraint~\eqref{eq:quan_l} is equivalently expressed as
\begin{align}
	\phi^{[\ell']}_n = \mathbf{b}_n^{[\ell']\mathsf{T}} \mathbf{q} \ \forall n \in \mathcal{N},
	\label{eq:bin_cons1}
\end{align}
where $\mathbf{b}_n^{[\ell']} \in \mathbb{B}^{{U} \times 1}$ is a binary selection vector. To ensure a unique selection, we impose
\begin{equation}
	\mathbf{b}_n^{[\ell'] \mathsf{T}} \mathbf{1}_{{U \times 1}} = 1 \ \forall n \in \mathcal{N}.
	\label{eq:bin_cons2}
\end{equation}
The above constraints can be compactly written as
\begin{subequations}
	\begin{align}
		& \boldsymbol{\phi}^{[\ell']}  = \mathbf{B}^{[\ell']} \mathbf{q}, \label{eq:Bin_con1} \\
		& \mathbf{1}_{{U \times 1}}^\mathsf{T} \mathbf{B}^{[\ell']}  = \mathbf{1}_{N}, \label{eq:Bin_con2}
	\end{align}
\end{subequations}
where $\mathbf{B}^{[\ell']} \triangleq [ \mathbf{b}_1^{[\ell']\mathsf{T}}, \ldots, \mathbf{b}_N^{[\ell']\mathsf{T}} ]^\mathsf{T}$. It should ne noted that

Hence, the resulting problem at iteration $t$ is
\begin{subequations}
	\begin{align}
		\max_{\boldsymbol{\Phi}^{[\ell']},\mathbf{B}^{[\ell']},\{\nu_{k,i} \}, \{\varrho_{k,i} \} } \quad &  \bar{R}_\text{sum}^{(t)} (\{ \nu_{k,i}\} , \{ \varrho_{k,i}\}) \\
		\text{s.t.} \ \ & \eqref{conp1}, \eqref{conp2}, \eqref{conp3}, \eqref{eq:Bin_con1}, \eqref{eq:Bin_con2}, \\
		& \mathbf{B}^{[\ell']} \in \mathbb{B}^{{N \times U}}.
	\end{align}
	\label{eq:Phi_FO}
\end{subequations}%
This problem is a mixed-integer program and can be efficiently solved using the CVX toolbox with the MOSEK solver~\cite{MOSEK}.

We summarize the proposed algorithm in \textbf{Algorithm~\ref{alg:main}}.

\begin{algorithm}[t]
	\caption{Sum rate maximization for the proposed SFIM- and DSIM-based systems.}
	\label{alg:main}
	
	\KwIn{ $\mathbf{W}^{(0)}$, $\hat{\mathbf{y}}^{(0)}$, $\boldsymbol{\phi}^{(0)}$, $t=0$, ${\bar{\varepsilon}} > 0$ }
	
	\Repeat{ $\big|R_{\mathrm{sum}}(\mathbf{W}^{(t)},\boldsymbol{\phi}^{(t)},\hat{\mathbf{y}}^{(t)})-R_{\mathrm{sum}}(\mathbf{W}^{(t-1)},\boldsymbol{\phi}^{(t-1)},\hat{\mathbf{y}}^{(t-1)})\big|\leq {\bar{\varepsilon}}$}{
		
		Update $t \leftarrow t+1$\;
		
		Update $\hat{\mathbf{y}}^{{(t)}}$ using \textbf{Algorithm~\ref{alg:y_update}}\;
		
		Update $\mathbf{W}^{{(t)}}$ by solving~\eqref{eq:W_FO}\;
		
		\For{$\ell' \in \mathcal{L}$}{
			Update $\boldsymbol{\phi}^{{[\ell'](t)}}$ by solving~\eqref{eq:Phi_FO}\;
		}
		
	}
	
	\KwOut{  $\mathbf{W} = \mathbf{W}^{{(t)}}$ , $\hat{\mathbf{y}} = \hat{\mathbf{y}}^{{(t)}}$, $\boldsymbol{\phi} = \boldsymbol{\phi}^{{(t)}}$ }
\end{algorithm}

\vspace{-0.3cm}
\subsection{Convergence Analysis}
The overall algorithm follows an AO framework, where each block of variables is updated while keeping the remaining blocks fixed. At each AO iteration, the corresponding subproblem is solved using either gradient projection, a penalty-based formulation, or SCA. Importantly, each update rule is constructed to generate a feasible iterate and to {yield a non-decreasing sequence of objective values for the original sum-rate maximization problem}. Since the objective function is {upper bounded} due to the transmit power budget constraint, which {induces compactness of the feasible set}, the resulting sequence of objective values is non-decreasing and convergent. Moreover, each subproblem is solved to a stationary point or satisfies the corresponding first-order optimality conditions, as ensured by the convergence properties of gradient projection methods for constrained optimization~\cite{bertsekas1999nonlinear,nocedal2006numerical}, penalty-based methods for the computation of Karush--Kuhn--Tucker (KKT) points~\cite{bertsekas1999nonlinear,nocedal2006numerical}, and SCA schemes for nonconvex optimization~\cite{marks1978general,scutari2017parallel}. Consequently, the sequence of AO iterates remains bounded, and every accumulation point of the generated sequence satisfies the block-wise first-order optimality conditions, thereby constituting a stationary (KKT) solution of the original nonconvex optimization problem.

\vspace{-0.5cm}
\section{Perturbation Analysis} \label{sec:per_ana}
This section presents a perturbation analysis for the SFIM and DSIM systems to gain insights into their achievable gains compared with the RSIM-based system. To keep the analysis tractable, we consider a SISO system with a single-antenna BS and a single receiving user. We further assume that the transmit power budget and the meta-atom response values are fixed. Moreover, the morphing displacement is assumed to be sufficiently small such that the distance constraints~\eqref{eq:dist_cons1} and~\eqref{eq:dist_cons2} are always satisfied.
\vspace{-0.2cm}

\subsection{SFIM}
The received signal power at the user can be expressed as
\begin{equation}
	P_\text{r} (\hat{\mathbf{y}})  = P_\text{t} \big| \mathbf{h}^\mathsf{T} (\hat{\mathbf{y}}) \boldsymbol{\Phi}^{[L]} \boldsymbol{\Omega}^{[L]} (\hat{\mathbf{y}}) \dots \boldsymbol{\Phi}^{[1]} \boldsymbol{\omega}^{[1]} (\hat{\mathbf{y}})  \big|^2,
\end{equation}
where $P_{\text{t}}$ denotes the transmit power, and $\mathbf{h}^{\mathsf{T}} (\hat{\mathbf{y}})$ and $\boldsymbol{\omega}^{[1]}(\hat{\mathbf{y}})$ represent the channel between the final SIM layer and the user, and the channel between the transmit antenna and the first SIM layer, respectively.

Using a first-order Taylor expansion of $P_\text{r}(\hat{\mathbf{y}})$ around $\hat{\mathbf{y}}=\mathbf{0}$, we obtain
\begin{equation}
	P^{\text{SFIM}}_\text{r} (\hat{\mathbf{y}})  \approx \underbrace{{P_\text{r} (\mathbf{0})}}_{P^{\text{RSIM}}_\text{r}} + \underbrace{{\hat{\mathbf{y}}^{\mathsf{T}} \nabla_{\hat{\mathbf{y}}}   P_\text{r} (\mathbf{0})}}_{G_\text{f} (\hat{\mathbf{y}})},
	\label{eq:PA_TA}
\end{equation}
where $P^{\text{RSIM}}_\text{r}$ denotes the received power of the RSIM-based system and $G_\text{f} (\hat{\mathbf{y}})$ represents the additional gain introduced by surface flexibility.

Assuming that the admissible morphing range lies within the region where the first-order approximation in~\eqref{eq:PA_TA} is accurate, we select the morphing vector as
\begin{equation}
	\hat{\mathbf{y}} = \tilde{y} \, \text{sign} \!\left(\nabla_{\hat{\mathbf{y}}}   P_\text{r} (\mathbf{0})\right).
\end{equation}
Accordingly, the flexibility gain achieved by the SFIM architecture can be approximated as
\begin{equation}
	G^{\text{SFIM}} \approx {\tilde{y}} \sum \nolimits_{\ell \in \mathcal{L}} \sum \nolimits_{n \in \mathcal{N}} \big| \partial P_\text{r} (\mathbf{0}) / \partial \hat{y}^{[\ell]}_n \big| = {\tilde{y}} \big\| \nabla_{\hat{\mathbf{y}}} P_\text{r} (\mathbf{0}) \big\|_1.
	\label{eq:gain_SFIM}
\end{equation}
It follows from~\eqref{eq:gain_SFIM} that the flexibility gain of the SFIM architecture scales approximately linearly with the morphing range, with a slope determined by the $\ell_1$-norm of the gradient of $P_\text{r}(\hat{\mathbf{y}})$ evaluated at $\hat{\mathbf{y}}=\mathbf{0}$. It should be emphasized, however, that the Taylor approximation in~\eqref{eq:PA_TA} is valid only locally. Consequently, the linear scaling behavior in~\eqref{eq:gain_SFIM} holds when the morphing range is sufficiently small. This observation is corroborated by the numerical results presented in Section~\ref{sec:sim_mor}, which also illustrate the behavior beyond the linear regime.

\subsection{DSIM}
Following an analysis analogous to that of the SFIM case, the flexibility-induced gain for the DSIM architecture can be expressed as
\begin{equation}
	G_\text{f} (\hat{\mathbf{y}}) = {\sum \nolimits_{\ell \in \mathcal{L}} \hat{y}^{[\ell]}} \sum \nolimits_{n \in \mathcal{N}} \partial P_\text{r} (\mathbf{0}) / \partial \hat{y}^{[\ell]}_n.
\end{equation}
Selecting the inter-layer displacements as
\begin{equation}
	\hat{y}^{[\ell]} = \tilde{y} \, \text{sign} \!\left( \sum \nolimits_{n \in \mathcal{N}} \partial P_\text{r} (\mathbf{0}) / \partial \hat{y}^{[\ell]}_n \right),
\end{equation}
the resulting DSIM flexibility gain becomes
\begin{equation}
	G^{\text{DSIM}} =  \tilde{y} \sum \nolimits_{\ell \in \mathcal{L}} \Big|\sum \nolimits_{n \in \mathcal{N}}  \partial P_\text{r} (\mathbf{0}) / \partial \hat{y}^{[\ell]}_n \Big| \leq G^{\text{SFIM}},
	\label{eq:gain_DSIM}
\end{equation}
where the inequality follows directly from the triangle inequality. Equation~\eqref{eq:gain_DSIM} indicates that the flexibility gain of the DSIM architecture also scales linearly with the morphing range, albeit with a smaller growth rate than that of SFIM due to the reduced number of degrees of control.

\section{Numerical Results} \label{sec:sim}

In this section, we present numerical results to examine the performance gains enabled by flexibility in SIM-enabled communications. We compare four system models: SFIM, DSIM and RSIM, which have already been discussed, and also hybrid SIM (HSIM), in which all layers are rigid except for the first and final layers, which are flexible.

Unless otherwise stated, the following parameters are used. The number of transmit antennas and the number of SIM layers are set to $M=6$ and $L=6$, respectively. Each SIM layer consists of $N=6\times6=36$ meta-atoms, and the number of users is $K=4$. The noise variance at each user is $\sigma_k^2=\unit[-104]{dBm}$. The carrier frequency is set to $f_\mathrm{c}=\unit[28]{GHz}$, which corresponds to a wavelength of $\lambda\approx\unit[10.7]{mm}$. The area of a transmit antenna and that of a meta-atom are both equal to $A_\mathrm{a}=A_\mathrm{m}=\frac{\lambda^2}{4}\approx28.6\ \unit{mm^2}.$ {The nominal inter-layer spacing between two consecutive SIM layers is set to $\bar{y}=6\lambda\approx\unit[64.2]{mm}$, which serves as a common reference geometry for all considered architectures.} The total transmit power budget is $P_\mathrm{max}=\unit[25]{dBm}$. {For flexible architectures, a minimum distance constraint between two meta-atoms in different layers is imposed as} $\epsilon=0.62\sqrt{\frac{\lambda^2}{4}}\approx\unit[3.32]{mm}.$ {The maximum allowable morphing displacement around the nominal layer position is set to} $\tilde{y}=\frac{\lambda}{2}\approx\unit[5.35]{mm},$ and the number of quantization bits is set to $2$, which correspond to 4 quantization levels.

{We assume that the distance between each user and the final SIM layer is} uniformly distributed between \unit[95]{m} and \unit[105]{m}, with both the azimuth and elevation angles uniformly distributed over the interval $[-\pi/4,\pi/4]$ rad. {Moreover, the propagation environment is modeled using $I=5$ scatterers}, each of which is assumed to have a distance uniformly distributed between \unit[55]{m} and \unit[105]{m}, with both the azimuth and elevation angles uniformly distributed over the interval $[-\pi/2,-\pi/4]$ rad. The average sum rate is evaluated via Monte Carlo simulations over 100 independent realizations of the user and scatterer locations. Finally, to ensure feasibility, the initial values of the meta-atom locations and responses are set to $\hat{\mathbf{y}}^{(0)}=\mathbf{0}_{NL\times1}$ and $\boldsymbol{\phi}^{(0)}=\mathbf{1}_{NL\times1}$, respectively, while the initial transmit beamformer $\mathbf{W}^{(0)}$ is obtained using zero-forcing. The rate threshold of the $k$-th user is then set to $R_k^{\mathrm{th}}=\log_{2}\big(1+0.95\gamma_{k}(\mathbf{W}^{(0)},\hat{\mathbf{y}}^{(0)},\boldsymbol{\phi}^{(0)})\big)$.

\subsection{Convergence}
\begin{figure}[tb]
	\begin{centering}
		\includegraphics[width=0.7\columnwidth]{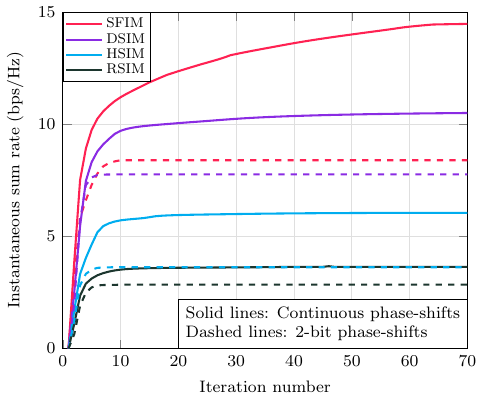}
		\par\end{centering}
	\caption{Sum-rate convergence of the proposed algorithm for different SIM architectures.}
	\label{fig:conv}
\end{figure}

\begin{figure*}
	\centering
	\subfloat[Layer 1.]{\includegraphics[width=0.27\textwidth]{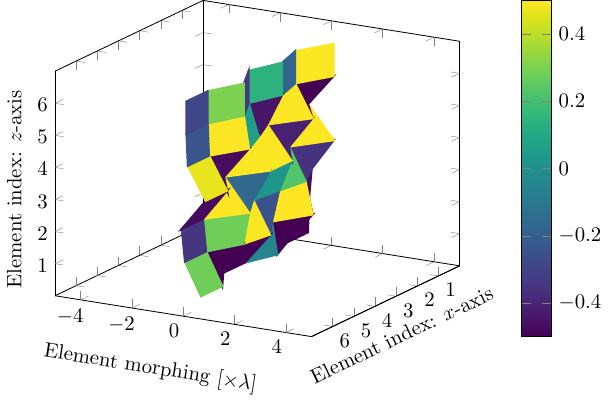}}%
	\hfill
	\subfloat[Layer 2.]{\includegraphics[width=0.27\textwidth]{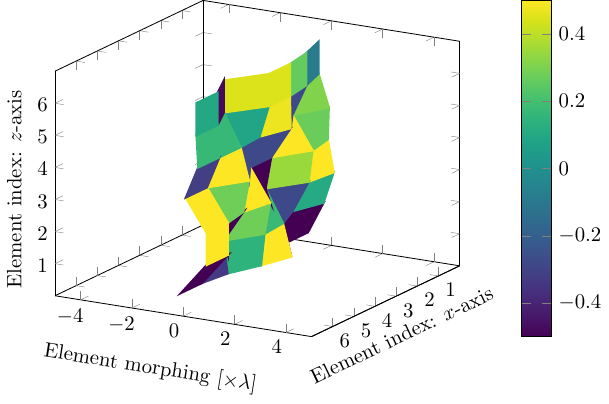}}%
	\hfill
	\subfloat[Layer 3.]{\includegraphics[width=0.27\textwidth]{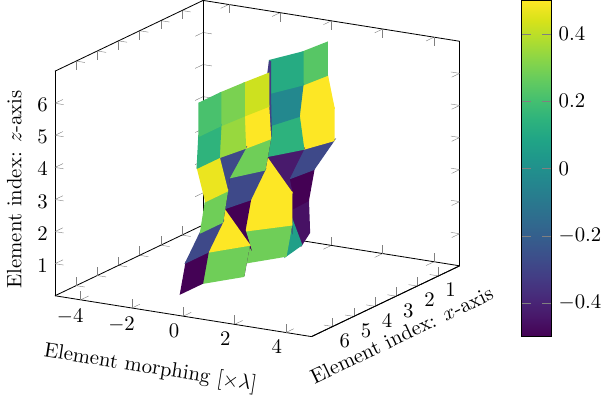}}\\[2mm]
	
	\subfloat[Layer 4.]{\includegraphics[width=0.27\textwidth]{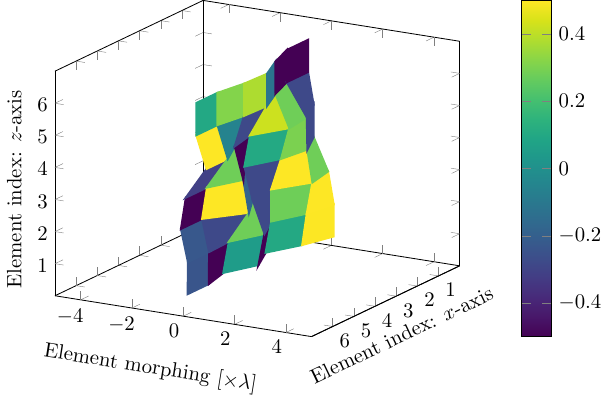}}%
	\hfill
	\subfloat[Layer 5.]{\includegraphics[width=0.27\textwidth]{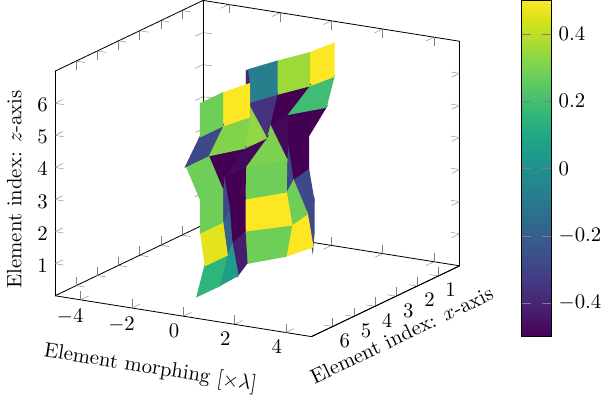}}%
	\hfill
	\subfloat[Layer 6.]{\includegraphics[width=0.27\textwidth]{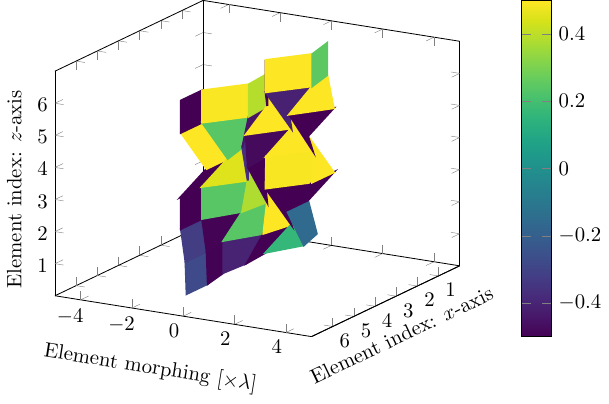}}
	
	\caption{Example optimized meta-atom morphing patterns of the SFIM layers (continuous phase case).}
    \vspace{-.5cm}
	\label{fig:SFIM_mor}
\end{figure*}

\begin{figure}[t]
	\begin{centering}
		\includegraphics[width=0.7\columnwidth]{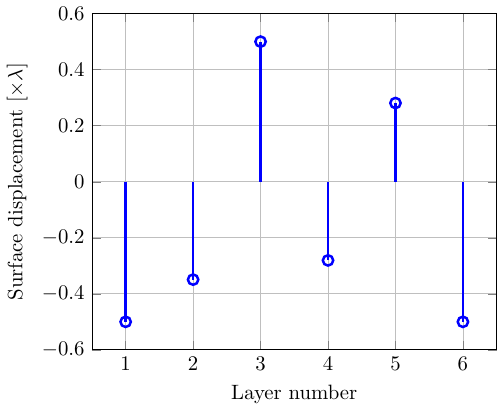}
		\par\end{centering}
	\caption{Example optimized inter-layer displacements in the DSIM architecture (continuous phase case).}
	\label{fig:DSIM_mor}
\end{figure}

Fig.~\ref{fig:conv} illustrates the convergence behavior of the sum rate versus the number of iterations. It can be observed that the sum rate for all architectures increases monotonically, with {most of the performance gains achieved within the first 10--20 iterations}. This {fast convergence behavior} indicates that the algorithm is well suited for practical scenarios requiring low-latency beamforming design. {Moreover, distinct convergence trends are observed between the continuous- and discrete-phase cases.} While the continuous-phase curves exhibit {gradual improvement even beyond 50 iterations}, the discrete-phase curves (dashed lines) {saturate much earlier}, typically stabilizing around iteration 10, {due to the restricted search space imposed by phase quantization}. Despite this early saturation, the discrete SFIM still maintains a significantly higher sum rate (by approximately $8.5$~bps/Hz) compared to the RSIM baseline.

Fig.~\ref{fig:SFIM_mor} visualizes the optimized physical configurations of the SFIM layers for the continuous meta-atom response case. It can be observed that the optimized surfaces exhibit {complex and non-trivial spatial variations}, indicating that the system effectively leverages its structural flexibility to perform {sophisticated wavefront shaping}. This behavior {extends beyond simple beam steering and contributes to enhanced sum-rate performance}. Moreover, the distinct topologies observed across the layers (from Layer~1 to Layer~6) {underscore the deep diffractive processing capability of the proposed architecture}, whereby the electromagnetic wave is successively refined as it propagates through the stacked layers. {These observations confirm that the proposed optimization framework successfully coordinates the cascaded layers to exploit the full spatial degrees of control of the SFIM structure for sum-rate maximization.}

Fig.~\ref{fig:DSIM_mor} depicts the optimized surface displacement for each layer of the DSIM architecture, illustrating how the system leverages {inter-layer positioning} as a degree of control. Unlike the complex spatial morphing observed in SFIM, the DSIM approach optimizes the global axial displacement of each layer. It can be observed that the resulting configuration is highly non-uniform, with displacements spanning the full range of $[-0.5 \lambda, 0.5 \lambda ]$. These pronounced deviations from the nominal uniform spacing indicate that the optimization algorithm actively exploits inter-layer spacing variations to enhance the sum rate.

\subsection{{Impact of the} Morphing Range} \label{sec:sim_mor}

\begin{figure*}
	\centering
	\begin{minipage}{.32\textwidth}
		\centering
		\includegraphics[width=1\columnwidth]{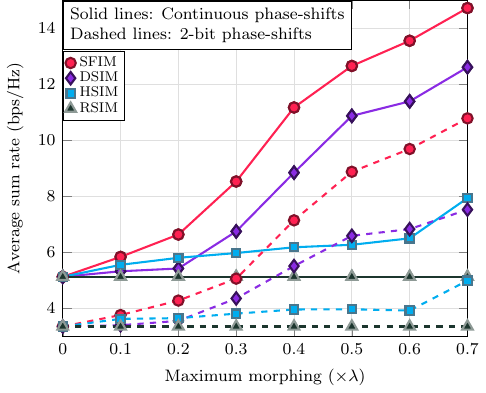}
		\caption{Average achievable sum rate under varying varying morphing ranges.}
		\label{fig:mor}
	\end{minipage}%
	\hfill 
	\begin{minipage}{.32\textwidth}
		\centering
		\includegraphics[width=\columnwidth]{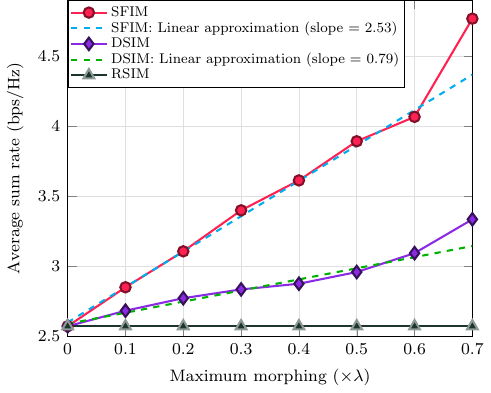}
		\caption{Average achievable sum rate under varying morphing ranges in the SISO scenario.}
		\label{fig:mor_SISO}
	\end{minipage}%
	\hfill 
	\begin{minipage}{.32\textwidth}
		\centering		
		\includegraphics[width=1\columnwidth]{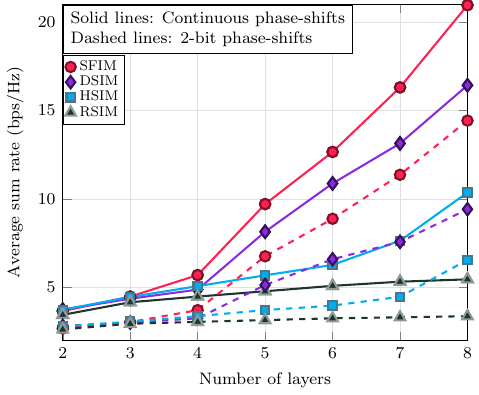}
		\caption{Average achievable sum rate with respect to the number of SIM layers.}
		\label{fig:num_lay}
	\end{minipage}
\end{figure*}

Fig.~\ref{fig:mor} shows the impact of the maximum allowable morphing range on the average sum rate. The results reveal a strong positive correlation between the flexibility range and the sum rate. Specifically, the SFIM architecture demonstrates a substantial performance improvement, increasing from approximately 5.2~bit/s/Hz in the rigid state (zero morphing) to over 14.5~bit/s/Hz when the morphing range extends to $0.7 \lambda$. Furthermore, a drastic change is observed between $0.3 \lambda$ and $0.5 \lambda$, where the performance slope is steepest. In contrast, the RSIM baseline remains static, while the HSIM shows only modest gains, highlighting that fully exploiting the morphing degree of control is essential for maximizing the sum rate in wave-based processing systems.


Fig.~\ref{fig:mor_SISO} depicts the average achievable sum rate for the SISO case (i.e., when $M=K=1$). In this scenario, we set $L=4$ to allow sufficient spacing between layers in order to investigate the linear gain discussed in Section~\ref{sec:per_ana}. It can be observed that, for both the SFIM and DSIM systems, a nearly linear increase in the sum rate can be realized for morphing ranges up to approximately $0.6\lambda$, beyond which the gain becomes even higher. This confirms that the linear approximation provides a conservative approximation of the achievable system performance.

Moreover, although the slope of the linear approximation is generally challenging to obtain, the resulting first-order model offers a practical means of estimating performance trends. For instance, Fig.~\ref{fig:mor_SISO} shows that increasing the maximum morphing range from $0.2\lambda$ to $0.5\lambda$ results in an increase from approximately 3.1~bps/Hz to 3.6~bps/Hz in the SFIM system, and from approximately 2.8~bps/Hz to 3.0~bps/Hz in the DSIM system. These gains can be predicted without reoptimizing the system, by directly evaluating the slope of the approximate linear model.

\subsection{Impact of the Number of Layers}
Fig.~\ref{fig:num_lay} depicts the average sum rate as a function of the number of SIM layers\footnote{As presented earlier, the total thickness of the SIM is fixed in this scenario, regardless of the number of layers.}. It can be observed that, as the number of layers increases, the gain achieved through surface flexibility increases significantly. For instance, with eight layers, SFIM, DSIM, and HSIM (with continuous meta-atom responses) achieve improvements of 284\%, 201\%, and 90\% in the sum rate compared to RSIM, respectively. In contrast, the RSIM architecture exhibits a clear performance saturation behavior, with the curve flattening after approximately five to six layers, indicating that propagation losses eventually limit the achievable wave-domain processing gains in the absence of flexibility. This limitation is effectively mitigated by exploiting flexibility in the SIM structure, as both SFIM and DSIM maintain a steep upward trajectory, {demonstrating that flexible architectures enable more effective utilization of additional layers for signal enhancement.} Furthermore, it is noteworthy that even the discrete versions of SFIM and DSIM significantly outperform the continuous RSIM when the number of layers exceeds five and seven, respectively, {highlighting the robustness and practical effectiveness of the proposed flexible architectures.}

\subsection{{Impact of} the Number of Meta-Atoms {per Layer}}

\begin{figure*}
	\centering
	\begin{minipage}{.32\textwidth}
		\centering
		\includegraphics[width=\columnwidth]{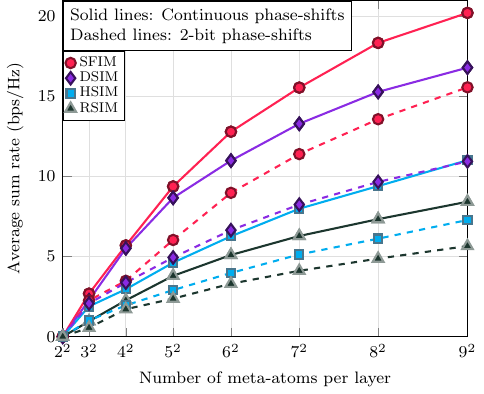}
		\caption{Average achievable sum rate under different numbers of meta-atoms per SIM layer.}
		\label{fig:num_meta}
	\end{minipage}%
	\hfill 
	\begin{minipage}{.32\textwidth}
		\centering
		\includegraphics[width=\columnwidth]{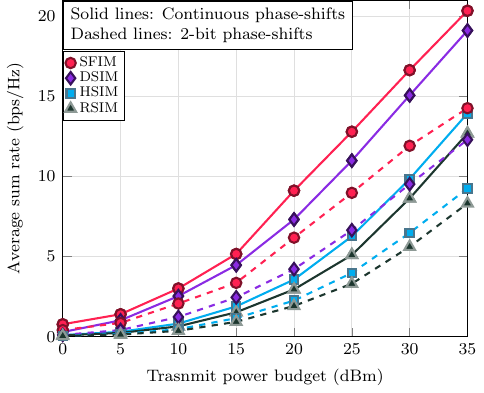}
		\caption{Average achievable sum rate under different transmit power budgets.}
		\label{fig:PPud}
	\end{minipage}%
	\hfill 
	\begin{minipage}{.32\textwidth}
		\centering		
		\includegraphics[width=\columnwidth]{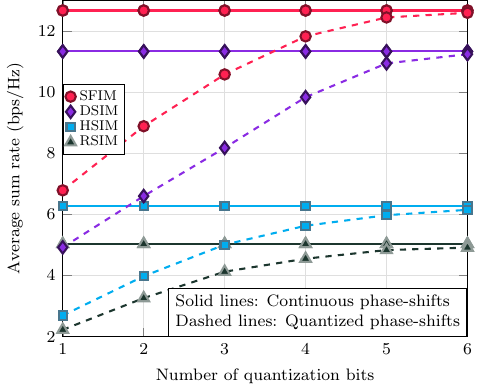}
		\caption{Average achievable sum rate under different meta-atom quantization resolutions.}
		\label{fig:QB}
	\end{minipage}
        \vspace{-.5cm}
\end{figure*}

Fig.~\ref{fig:num_meta} depicts the average sum rate as a function of the number of meta-atoms per layer, highlighting the relationship between aperture size and spectral efficiency. It can be observed that the sum rate increases monotonically for all architectures as the number of meta-atoms grows, confirming that a larger aperture provides finer beamforming granularity and improved spatial resolution; however, the magnitude of this gain varies significantly across configurations. The flexible SFIM and DSIM architectures exhibit a steep performance trajectory, {reflecting their ability to effectively exploit the increased spatial degrees of control} to enhance signal focusing and interference mitigation. For instance, as the number of meta-atoms increases to 81, the continuous SFIM achieves a sum rate exceeding 20~bit/s/Hz, whereas the RSIM reaches only roughly 8.5~bit/s/Hz. Moreover, the superiority of flexible architectures is preserved under discrete phase constraints: although the quantization penalty (i.e., the gap between solid and dashed curves) generally widens as the system scales, the discrete versions of SFIM and DSIM consistently outperform the continuous RSIM baseline. Specifically, with 81 meta-atoms, the discrete SFIM achieves a sum rate of approximately 15.5~bit/s/Hz, {corresponding to a gain of nearly 82\%} over the continuous RSIM.

\subsection{{Impact of} the Transmit Power Budget}
Fig.~\ref{fig:PPud} illustrates the average achievable sum rate as a function of the transmit power budget. It can be observed that flexible SIM architectures achieve the same target sum rate with a significantly lower power budget compared to RSIM. For example, to achieve an average sum rate of approximately $9$~bit/s/Hz in the discrete case, SFIM requires about $25$~dBm, whereas RSIM requires more than $35$~dBm, corresponding to a power saving of at least $10$~dB. Similarly, DSIM and HSIM achieve a sum rate of around $6$~bit/s/Hz at approximately $20$~dBm, while RSIM requires nearly $30$~dBm to reach a comparable performance. {These results indicate that structural flexibility yields substantial power savings by enabling more efficient beamforming and signal focusing, even under discrete meta-atom response constraints.} Moreover, {the robustness of this gain to hardware quantization is evident from the fact that the discrete SFIM configuration consistently outperforms the continuous RSIM baseline}, indicating that the beamforming gains obtained via flexible-layer optimization are sufficient to compensate for, and even exceed, the performance degradation induced by discrete phase shifts in rigid architectures.

\subsection{{Impact of} the Number of Quantization Bits}
Fig.~\ref{fig:QB} shows the impact of meta-atom response quantization on system performance. It is observed that the sum rate of the discrete systems increases monotonically with the number of quantization bits, rapidly converging toward the continuous upper bounds. Specifically, while a 1-bit resolution results in a substantial performance degradation, reducing the SFIM sum rate by nearly 46\% compared to the continuous case, increasing the resolution to 4 bits {largely mitigates this loss}. For instance, with 4 quantization bits, the discrete SFIM and DSIM architectures achieve approximately 93\% and 95\% of their theoretical maxima, respectively, {indicating that a moderate quantization resolution is sufficient to closely approach continuous-phase performance.}

\section{Conclusion} \label{sec:conc}
This paper examined the performance improvements enabled by incorporating structural flexibility into SIM-assisted communication systems. Two flexible SIM architectures were considered: DSIM, in which the inter-layer spacing was optimized, and SFIM, where each layer was fully morphable. A joint optimization framework was developed to design the meta-atom locations and responses together with the transmit beamformer, with the objective of maximizing the system sum rate under user rate, quantization, morphing, and inter-layer distance constraints. {To address the resulting non-convex design problem}, an alternating optimization strategy was proposed, integrating gradient projection, penalty-based techniques, and successive convex approximation. Furthermore, a perturbation analysis was conducted to provide insights into the benefits of flexibility, revealing that the achievable gains of both SFIM and DSIM {scale approximately linearly} with the morphing range, with SFIM exhibiting a steeper growth rate. Simulation results confirmed that the proposed flexible SIM architectures effectively alleviate performance saturation as the number of layers increases, while also achieving significant power savings compared to conventional rigid SIM designs.

\ifCLASSOPTIONcaptionsoff
  \newpage
\fi





\appendices


\bibliographystyle{IEEEtran}
\bibliography{IEEEabrv,Bibliography}
%


\end{document}